\documentclass[twocolumn,aps,prc,floatfix,showpacs,superscriptaddress]{revtex4}
\usepackage{graphicx}
\usepackage{amsmath}
\usepackage{amssymb}
\usepackage{indentfirst}	%added by RMM 
\usepackage{float}		%added by RMM 
\usepackage{afterpage}	%added by RMM 

\newcommand{\sss}{\scriptscriptstyle}

\newcommand{\peek}{{p(e,e'K^+)}}
\newcommand{\betatof}{\beta_{\mathrm{tof}}}

\newcommand{\betak}{\beta_{\mathrm{\sss{K}}}}
\newcommand{\Sigo}{\Sigma^{0}}
\newcommand{\sigt}{\sigma_{\sss{T}}}
\newcommand{\sigl}{\sigma_{\sss{L}}}

\newcommand{\thcm}{\theta_{\sss{qK}}^{*}}
\newcommand{\avgthcm}{\thcm}

\begin{document}

\title{Separation of the Longitudinal and Transverse Cross Sections \\
in the $\peek\Lambda$ and $\peek\Sigo$ Reactions}

\author{R.~M.~Mohring}
\affiliation{University of Maryland, College Park, MD 20742, USA}
\author{D.~Abbott}
\affiliation{Thomas Jefferson National Accelerator Facility, Newport News, VA 23606, USA}
\author{A.~Ahmidouch}
\affiliation{Kent State University,~Kent,~OH~44242,~USA}
\author{Ts.~A.~Amatuni}
\affiliation{Yerevan Physics Institute,~Yerevan,~Armenia}
\author{P.~Ambrozewicz}
\affiliation{Temple University, Philadelphia, PA 19122, USA}
\author{T.~Angelescu}
\affiliation{Bucharest University, Bucharest, Romania}
\author{C.~S.~Armstrong}
\affiliation{College of William and Mary, Williamsburg, VA 23187, USA}
\author{J.~Arrington}
\affiliation{Argonne National Laboratory, Argonne, IL 60439, USA}
\author{K.~Assamagan}
\affiliation{Hampton University, Hampton, VA 23668, USA}
\author{S.~Avery}
\affiliation{Hampton University, Hampton, VA 23668, USA}
\author{K.~Bailey}
\affiliation{Argonne National Laboratory, Argonne, IL 60439, USA}
\author{K.~Beard} 
\affiliation{Hampton University, Hampton, VA 23668, USA}
\author{S.~Beedoe} 
\affiliation{North Carolina A \& T State University, Greensboro, NC 27411, USA}
\author{E.~J.~Beise} 
\affiliation{University of Maryland, College Park, MD 20742, USA}
\author{H.~Breuer}
\affiliation{University of Maryland, College Park, MD 20742, USA}
\author{R.~Carlini} 
\affiliation{Thomas Jefferson National Accelerator Facility, Newport News, VA 23606, USA}
\author{J.~Cha} 
\affiliation{Hampton University, Hampton, VA 23668, USA}
\author{C.~C.~Chang} 
\affiliation{University of Maryland, College Park, MD 20742, USA}
\author{N.~Chant}
\affiliation{University of Maryland, College Park, MD 20742, USA}
\author{E.~Cisbani} 
\affiliation{Physics Laboratory, Istituto Superiore di Sanit\'{a} and INFN-Sezione Sanit\'{a}, Rome, Italy}
\author{G.~Collins} 
\affiliation{University of Maryland, College Park, MD 20742, USA}
\author{W.~Cummings}
\affiliation{Argonne National Laboratory, Argonne, IL 60439, USA}
\author{S.~Danagoulian} 
\affiliation{North Carolina A \& T State University, Greensboro, NC 27411, USA}
\affiliation{Thomas Jefferson National Accelerator Facility, Newport News, VA 23606, USA}
\author{R.~De Leo} 
\affiliation{Physics Laboratory, Istituto Superiore di Sanit\'{a} and INFN-Sezione Sanit\'{a}, Rome, Italy}
\author{F.~Duncan}
\affiliation{University of Maryland, College Park, MD 20742, USA}
\author{J.~Dunne} 
\affiliation{Thomas Jefferson National Accelerator Facility, Newport News, VA 23606, USA}
\author{D.~Dutta} 
\affiliation{Northwestern University, Evanston, IL 60201, USA}
\author{T.~Eden} 
\affiliation{Hampton University, Hampton, VA 23668, USA}
\author{R.~Ent}
\affiliation{Thomas Jefferson National Accelerator Facility, Newport News, VA 23606, USA}
\author{L.~Eyraud} 
\affiliation{ISN, IN2P3-UJF, 38026 Grenoble, France}
\author{L.~Ewell}
\affiliation{University of Maryland, College Park, MD 20742, USA}
\author{M.~Finn} 
\affiliation{College of William and Mary, Williamsburg, VA 23187, USA}
\author{T.~Fortune}
\affiliation{University of Pennsylvania, Philadelphia, PA 19104, USA}
\author{V.~Frolov} 
\affiliation{Rensselaer Polytechnic Institute, Troy, NY 12180, USA}
\author{S.~Frullani} 
\affiliation{Physics Laboratory, Istituto Superiore di Sanit\'{a} and INFN-Sezione Sanit\'{a}, Rome, Italy}
\author{C.~Furget}
\affiliation{ISN, IN2P3-UJF, 38026 Grenoble, France}
\author{F.~Garibaldi} 
\affiliation{Physics Laboratory, Istituto Superiore di Sanit\'{a} and INFN-Sezione Sanit\'{a}, Rome, Italy}
\author{D.~Gaskell} 
\affiliation{Oregon State University, Corvallis, OR 97331, USA}
\author{D.~F.~Geesaman}
\affiliation{Argonne National Laboratory, Argonne, IL 60439, USA}
\author{P.~Gueye}
\affiliation{Hampton University, Hampton, VA 23668, USA}
\author{K.~K.~Gustafsson}
\affiliation{University of Maryland, College Park, MD 20742, USA}
\author{J.~-O.~Hansen} 
\affiliation{Argonne National Laboratory, Argonne, IL 60439, USA}
\author{M.~Harvey}
\affiliation{Hampton University, Hampton, VA 23668, USA}
\author{W.~Hinton} 
\affiliation{Hampton University, Hampton, VA 23668, USA}
\author{E.~Hungerford} 
\affiliation{University of Houston, Houston, TX 77204, USA}
\author{M.~Iodice}
\affiliation{Physics Laboratory, Istituto Superiore di Sanit\'{a} and INFN-Sezione Sanit\'{a}, Rome, Italy}
\author{C.~Jackson} 
\affiliation{North Carolina A \& T State University, Greensboro, NC 27411, USA}
\author{C.~Keppel} 
\affiliation{Hampton University, Hampton, VA 23668, USA}
\affiliation{Thomas Jefferson National Accelerator Facility, Newport News, VA 23606, USA}
\author{W.~Kim}
\affiliation{Kyungpook National University, Taegu, S.~Korea}
\author{K.~Kino}
\affiliation{Tohoku University,~Sendai~982,~Japan} 
\author{D.~Koltenuk} 
\affiliation{University of Pennsylvania, Philadelphia, PA 19104, USA}
\author{S.~Kox}
\affiliation{ISN, IN2P3-UJF, 38026 Grenoble, France}
\author{L.~Kramer}
\affiliation{Florida International University, Miami, FL 33199, USA }
\affiliation{Thomas Jefferson National Accelerator Facility, Newport News, VA 23606, USA}
\author{T.~Leone} 
\affiliation{INFN, Sezione di Lecce, via Arnesano, 73100 Lecce, Italy}
\author{A.~Lung}
\affiliation{University of Maryland, College Park, MD 20742, USA}
\author{D.~Mack} 
\affiliation{Thomas Jefferson National Accelerator Facility, Newport News, VA 23606, USA}
\author{R.~Madey} 
\affiliation{Hampton University, Hampton, VA 23668, USA}
\affiliation{Kent State University,~Kent,~OH~44242,~USA}
\author{M.~Maeda} 
\affiliation{Tohoku University,~Sendai~982,~Japan} 
\author{S.~Majewski}
\affiliation{Thomas Jefferson National Accelerator Facility, Newport News, VA 23606, USA}
\author{P.~Markowitz} 
\affiliation{Florida International University, Miami, FL 33199, USA }
\affiliation{Thomas Jefferson National Accelerator Facility, Newport News, VA 23606, USA}
\author{T.~Mart}
\affiliation{Jurusan Fisika, FMIPA, Universitas Indonesia, Depok 16424, Indonesia}
\author{C.~J.~Martoff}
\affiliation{Temple University, Philadelphia, PA 19122, USA}
\author{D.~Meekins}
\affiliation{College of William and Mary, Williamsburg, VA 23187, USA}
\author{A.~Mihul}
\affiliation{Bucharest University, Bucharest, Romania}
\author{J.~Mitchell}
\affiliation{Thomas Jefferson National Accelerator Facility, Newport News, VA 23606, USA}
\author{H.~Mkrtchyan}
\affiliation{Yerevan Physics Institute,~Yerevan,~Armenia}
\author{S.~Mtingwa}
\affiliation{North Carolina A \& T State University, Greensboro, NC 27411, USA}
%\author{G.~Niculescu}
%\affiliation{Hampton University, Hampton, VA 23668, USA}
\author{I.~Niculescu}
\affiliation{Hampton University, Hampton, VA 23668, USA}
\author{R.~Perrino} 
\affiliation{INFN, Sezione di Lecce, via Arnesano, 73100 Lecce, Italy}
\author{D.~Potterveld} 
\affiliation{Argonne National Laboratory, Argonne, IL 60439, USA}
\author{J.~W.~Price}
\affiliation{Rensselaer Polytechnic Institute, Troy, NY 12180, USA}
\author{B.~A.~Raue} 
\affiliation{Florida International University, Miami, FL 33199, USA }
\affiliation{Thomas Jefferson National Accelerator Facility, Newport News, VA 23606, USA}
\author{J.-S.~Real} 
\affiliation{ISN, IN2P3-UJF, 38026 Grenoble, France}
\author{J.~Reinhold}
\affiliation{Argonne National Laboratory, Argonne, IL 60439, USA}
\author{P.~Roos} 
\affiliation{University of Maryland, College Park, MD 20742, USA}
\author{T.~Saito}
\affiliation{Tohoku University,~Sendai~982,~Japan}
\author{G.~Savage} 
\affiliation{Hampton University, Hampton, VA 23668, USA}
\author{R.~Sawafta}
\affiliation{North Carolina A \& T State University, Greensboro, NC 27411, USA}
\affiliation{Thomas Jefferson National Accelerator Facility, Newport News, VA 23606, USA}
\author{R.~Segel} 
\affiliation{Northwestern University, Evanston, IL 60201, USA}
\author{S.~Stepanyan} 
\affiliation{Yerevan Physics Institute,~Yerevan,~Armenia}
\author{P.~Stoler}
\affiliation{Rensselaer Polytechnic Institute, Troy, NY 12180, USA}
\author{V.~Tadevosian} 
\affiliation{Yerevan Physics Institute,~Yerevan,~Armenia}
\author{L.~Tang}
\affiliation{Hampton University, Hampton, VA 23668, USA}
\affiliation{Thomas Jefferson National Accelerator Facility, Newport News, VA 23606, USA}
\author{L.~Teodorescu}
\affiliation{Bucharest University, Bucharest, Romania}
\author{T.~Terasawa} 
\affiliation{Tohoku University,~Sendai~982,~Japan}
\author{H.~Tsubota} 
\affiliation{Tohoku University,~Sendai~982,~Japan}
\author{G.~M.~Urciuoli}
\affiliation{Physics Laboratory, Istituto Superiore di Sanit\'{a} and INFN-Sezione Sanit\'{a}, Rome, Italy}
\author{J.~Volmer} 
\affiliation{Free University,~Amsterdam,~The~Netherlands}
\author{W.~Vulcan} 
\affiliation{Thomas Jefferson National Accelerator Facility, Newport News, VA 23606, USA}
\author{T.~P.~Welch} 
\affiliation{Oregon State University, Corvallis, OR 97331, USA}
\author{R.~Williams}
\affiliation{Hampton University, Hampton, VA 23668, USA}
\author{S.~Wood} 
\affiliation{Thomas Jefferson National Accelerator Facility, Newport News, VA 23606, USA}
\author{C.~Yan} 
\affiliation{Thomas Jefferson National Accelerator Facility, Newport News, VA 23606, USA}
\author{B.~Zeidman}
\affiliation{Argonne National Laboratory, Argonne, IL 60439, USA}

%%% These are the people who have asked to be removed for one reason
%%% or another.
%%%O.~K.~Baker$^{2,3}$, asked to be removed...
%%%G.~Lolos$^{12}$,   never worked on the exp.

\date{\today}

\begin{abstract}
We report measurements of cross sections for the
reaction $p$($e$,$e^\prime K^+$)Y, for both the $\Lambda$ and
$\Sigo$ hyperon states, at an invariant mass of $W$=1.84~GeV
and four-momentum
transfers $0.5 <  Q^2 < 2 $ (GeV/c)$^2$. Data were taken
for three values of virtual photon
polarization $\epsilon$, allowing the decomposition of the
cross sections into longitudinal and transverse components.
The $\Lambda$ data is a revised analysis of prior work, whereas the
$\Sigo$ results have not been previously reported.
\end{abstract}

%\keywords{}

\maketitle

%\narrowtext

%Include the various tex documents/sections here: For PRC submission this has
% be one big file.

%\input{intro}
%\input{apparatus}
%\input{dataanalysis}
%\input{results}
%\input{references}

\section{Introduction}
\label{sec:intro}

Electromagnetic production of strange baryons has long been of
interest because of the fact that such data can provide unique
information about the flavor dependence of nucleon excited states,
eventually leading to a better understanding of the theory of QCD.
Unfortunately, progress in understanding the production mechanism
has been slow, in no small part because of the lack of high
quality data.  With the recent availability of a high quality,
continuous electron beam at Jefferson Laboratory, precise new
measurements are now achievable over a wide kinematic
range~\cite{gnprl,JLABK}. In addition to providing information
on the elementary production reaction, such data will also be
a benchmark for future investigations of hyperon-nucleon
interactions with nuclear targets.

In this paper, we present new cross section data on the reaction
$p(e,e^\prime K^+)\Sigma^0$ from Jefferson Lab experiment E93-018,
that were acquired at values of the square of four-momentum
transfer, $Q^2$, between 0.5
and 2.0 (GeV/c)$^2$.  At each value of $Q^2$, cross sections were
obtained at three different
values of the virtual photon polarization $\epsilon$,
allowing a separation of the cross section into its longitudinal (L)
and transverse (T) components.  Results for the
$\Lambda$ channel were previously reported in Ref.~\cite{gnprl}. 
However, in order to
provide an internally consistent comparison between the two
reaction channels, we also present a reanalysis of the
$\Lambda$ data. The differences between the two analyses
will be discussed in detail in section~\ref{sec:mc}:
here we use a direct comparison of the experimental data to
simulated yields in order to extract the cross section. The 
result is a significantly smaller longitudinal contribution and 
a weaker $Q^2$ dependence than was previously reported. 
After a brief introduction and a detailed description of the experiment
and analysis, the data will be compared to isobaric models of
meson electroproduction described below.

Exploratory measurements of kaon electroproduction were first
carried out between 1972 and 1979~\cite{bebek-lt,brauel}.  In
Ref.~\cite{bebek-lt} the longitudinal and transverse contributions to
the $\Lambda$ and $\Sigo$ cross sections were separated for three
values of $Q^2$. Two values of $\epsilon$ were measured for each
kinematic setting with relatively poor statistical precision, and
the uncertainties in the L/T separated results were large. In
Ref.~\cite{brauel}, the measurements were focused on pion
electroproduction, but a sample of kaons was also acquired, from
which cross sections were extracted. These measurements provided
the first determination of the qualitative behavior of the kaon
cross sections and were the basis for the development of modern
models of kaon electroproduction.

Theoretical models all attempt to reproduce the available data
from both kaon production and radiative kaon capture, while
maintaining consistency with SU(3) symmetry constraints on the
coupling constants \cite{wjc1992,cwj1993}. The energy regime addressed by
these models is low enough that they are formulated using
meson-nucleon degrees of freedom. 
The most recent theoretical
efforts can be divided into two categories, isobaric models and
those that use Regge trajectories.

\begin{figure}
\begin{center}
\includegraphics[width=3.3in]{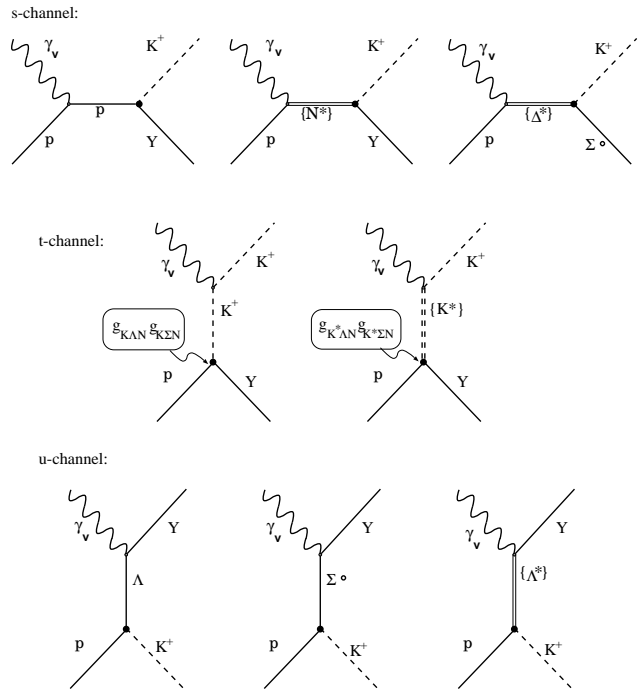}
\caption{Examples of Feynman diagrams for kaon electroproduction
considered in the isobaric models.  The couplings in the
$t$-channel ($g_{\sss K\Lambda N}$, etc...) are shown explicitly.
Note that the $s$-channel processes involving $\Delta^*$
resonances are forbidden by isospin conservation for $\Lambda$
production.}
\label{fig:kdiagrams}
\end{center}
\end{figure}

The approach taken in the isobaric models is to explicitly
calculate kaon production amplitudes from tree-level (\textit{i.e.},
only one particle exchanged) processes.  A typical
selection of diagrams considered is depicted in
Figure~\ref{fig:kdiagrams}.  For example, in
David~\textit{et~al.}~\cite{saclay-lyon}, the authors
sum over $s$-channel nucleon resonances up to and including
spin~5/2, $u$-channel hyperon resonances of spin~1/2, and
$t$-channel kaon resonances K$^*$(892) and K1(1270).  In the WJC
model~\cite{wjc1992}, a different selection of $s$ and $u$ channel
resonances is included (both limited to spin 3/2).  
Mart~\textit{et al.}~\cite{Mar00}
include the lowest lying $S$- and $P$-wave resonances, plus
an additional resonance $D_{13}$ for which there appears to
be evidence from kaon photoproduction~\cite{Tra98}, although alternative
interpretations of the data have been put forth in~\cite{Sag01} 
and~\cite{Jan02}.

The various isobaric models share the
property that they initially include
only spin~1/2 baryonic resonances (although the specific
resonances differ) and determine the remaining coupling constants
from performing phenomenological fits to the data. The coupling
constants, which are the parameters of the theory, are not
well-constrained due to the lack of available data.  Because of
this,  one sees differences in the various models for quantities
such as $g_{\sss K\Lambda N}$. One issue that these differences
reflect is that the various models disagree on the relative
importance of the resonances entering the calculation.  Further
details on the various isobaric models can be found in
\cite{wjc1992,saclay-lyon,adelseck-saghai,mbh}.

% === GRAPHICS ====================================
% ==============================================

Models based on Regge trajectories~\cite{reggetheory} were
developed in the early 1970's to describe pion photoproduction
data, of which there is a relative abundance. This approach has
recently been revisited by Vanderhaegen, Guidal, and Laget (VGL)
\cite{vgl-electroproduction}. Here, the standard single particle
Feynman propagator, $1/(t-m^2)$, is replaced by a Regge propagator
that accounts for the exchange of a family of particles with the
same internal quantum numbers \cite{vgl1}.  The extension of the
photoproduction model to electroproduction is accomplished by
multiplying the gauge invariant $t$-channel K and K$^*$ diagrams
by a form factor (the isobaric models such as WJC also include
electromagnetic form factors, however the functional forms used
differ between models \cite{wjc1992,adelseck-saghai}). For the VGL
model this is given as a monopole form factor,
\begin{equation}
F_{\sss K,K^*}(Q^2) = \frac{1}{1+Q^2/\Lambda^2_{\sss K, K^*}} \, \, ,
\end{equation}
where $\Lambda^2_{\sss K, K^*}$ are mass scales that are
essentially free parameters, but can be fixed to fit the high
$Q^2$ behavior of the separated electroproduction cross sections,
$\sigt$ and $\sigl$. For both the isobaric and Regge approaches,
precise experimental results for longitudinal/transverse separated
cross sections are important for placing constraints on the free
parameters within the models, hopefully giving insight into
the reaction mechanisms.

An additional motivation for performing measurements of L/T separated
cross sections in kaon electroproduction is to determine the $Q^2$ 
dependence of the K$^+$ electromagnetic form factor. 
If it can be demonstrated that $\sigma_L$ is dominated by photon absorption
on a ground state kaon, the K$^+$ form factor can be extracted
through a measurement of the $t$-dependence of the 
longitudinal component of the cross section.  This technique has been
used to determine the $\pi^+$ electromagnetic form factor,
including a recent new measurement~\cite{Vol01}. While the kaon
form factor may not be able to be extracted from the data presented here, 
it is the subject of other recently completed  
measurements at Jefferson Laboratory~\cite{Mar01}.

Finally, historically there has been  
interest in the ratio of $\Sigo/\Lambda$ transverse cross
sections, which is linked to contribution of sea quarks to nucleon
structure~\cite{Cley75,Nach74}. Within the context of the parton
model, isospin arguments would lead one to expect
the $\Sigo/\Lambda$ transverse cross section ratio
at forward kaon CM angles to approach 0 with increasing 
Bjorken $x$ if the kaon production mechanism is dominated by the 
photon interacting with single valence $u$ quark. When sea
quarks are included in the nucleon's initial state, the approach
to 0 is expected to be more rapid. Measurement of
of $\sigt(\Sigo)/\sigt(\Lambda)$ over a broad range of $x$
may provide information about the intrinsic $\overline{s}$$s$ content 
of the proton.

%%%%%%%

\subsection{Elementary Kaon Electroproduction}
\begin{figure}
\begin{center}
\includegraphics[width=3.3in]{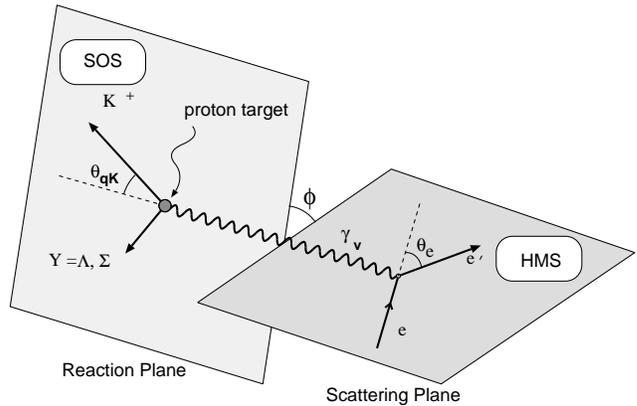}
\caption{Definition of the kaon electroproduction reaction. Note the azimuthal
angle, $\phi$, between the scattering and reaction planes with respect to the
direction of the virtual photon.}
\label{fig:peekdiag}
\end{center}
\end{figure}

The elementary kaon electroproduction reaction studied here,
\begin{center}
$e$ + $p$ $\rightarrow$ $e'$ + $K^+$ + ($\Lambda$ or $\Sigo$) ,
\end{center}
is shown in Figure \ref{fig:peekdiag}.  An incident electron
($e$) with lab energy $E$, scatters by radiating a 
virtual photon ($\gamma_{v}$). The
scattered electron ($e'$) travels at a polar angle $\theta_e$ in
the Laboratory frame with respect to the direction of the incident
beam, defining the scattering plane. The virtual photon carries
momentum $\vec{q}$ and energy $\nu$  and interacts with a target
proton to form a kaon (K$^+$) and a hyperon (Y, here either a
$\Lambda$ or $\Sigo$). The kaon travels at a polar angle
$\theta_{\sss{qK}}$ in the Laboratory frame with respect to the
virtual photon direction and is also detected. The reaction plane,
defined by the produced kaon and produced hyperon, makes an angle
$\phi$  with respect to the scattering plane.

% === GRAPHICS ====================================
% ==============================================

 The exclusive five-fold
differential electroproduction cross section can be expressed in
terms of a virtual photoproduction cross section,
$\frac{d^{\,2}\sigma}{d\Omega_{\sss K}^*}$, multiplied by a virtual
photon flux factor, $\Gamma_{0}$~\cite{devlyth}. The cross
section is written in
terms of the scattered electron energy, $E'$, electron Lab frame
solid angle, $d\Omega'_{e} \equiv d\cos\theta_e \, d\phi_e \,$,
and kaon Center-of-Momentum frame (CM) solid angle, $d\Omega_{\sss
K}^{*} \equiv d\cos\theta_{\sss qK}^* \, d\phi \,$ as
\begin{equation}
\label{eq:virtphot} 
\frac{d^{\,5}\sigma}{dE' d\Omega'_{e}
d\Omega_{\sss K}^{*}} = \Gamma_{0}(E', \Omega'_{e}) \,
\left(\frac{d^{\,2}\sigma}{d\Omega_{\sss K}^{*}}\right) \, \, ,
\end{equation}
with
\begin{equation}
\Gamma_{0}(E', \Omega'_{e}) = \frac{\alpha}{2\pi^{2}} \, 
\frac{(W^{2}-m_p^{2})}{2 \, m_p} \, \frac{E'}{E} \, \frac{1}{Q^{2}} \,
\frac{1}{(1 - \epsilon)} \, .
\end{equation}

Here the CM frame is that of the (virtual photon + proton)
system, and that the CM frame counterparts to Lab variables will
be denoted with an asterisk superscript. In this expression,
$\alpha$ is the fine structure constant ($\approx$1/137), $m_p$ is
the proton mass, $W$ is the total energy of the (virtual photon
+ proton) system, $Q^2$ is the square of the four-momentum
transfer carried by the virtual photon, and $\epsilon$ is the
polarization of the virtual photon, given by
\begin{equation}
\label{eq:epsilon}
\epsilon = \frac{1}{1 + 2 \, \frac{|\mathbf{q}|^{2}}{Q^{2}} \, 
\tan^2\frac{\theta_e}{2} } \, \, ,
\end{equation}
where $\mathbf{q}$ is the virtual photon three-momentum.

Because the beam and target were unpolarized, and no
outgoing polarization was measured, the virtual photoproduction
cross section can be decomposed into four terms:
\begin{equation}
\label{eq:ltxsect} 
\frac{d^{\,2}\sigma}{d\Omega_{\sss K}^{*}} =
\sigma_{\sss T} + \epsilon \: \sigma_{\sss L} +
\sqrt{2\epsilon(\epsilon + 1)} \: \sigma_{\sss LT} \cos \phi  +
\epsilon \: \sigma_{\sss TT} \cos 2\phi \, \, ,
\end{equation}
where
 $\sigma_{\sss T}$~is the cross section due to transversely polarized
 virtual photons, $\sigma_{\sss L}$~is due to longitudinally polarized
 virtual photons, and $\sigma_{\sss LT}$ and $\sigma_{\sss TT}$ are
 interference terms between two different polarization states.
 If one integrates the cross section over all $\phi
\in (0,2\pi)$, the interference terms vanish leaving only the
combined contributions from the transverse and longitudinal cross
sections, $\sigma_{\sss T} + \epsilon \sigma_{\sss L}$. By
measuring the cross section at several values of the virtual
photon polarization, $\epsilon$, the cross sections $\sigma_{\sss
T}$ and $\sigma_{\sss L}$ can be separated. The experimental setup
described here was such that at each of the four values of $Q^2$,
the full range in $\phi \in (0,2\pi)$ was accessible at three
different values of the virtual photon polarization, $\epsilon$,
as listed in Table \ref{tab:kinematics}. The data were fitted
using the linear dependence between  ($\sigma_{\sss T} + \epsilon
\sigma_{\sss L}$) and $\epsilon$. The intercept and slope of the
fitted line were used to extract $\sigma_{\sss T}$ and
$\sigma_{\sss L}$ for each $Q^2$, for both the $\Lambda$ and
$\Sigma^0$ channels.

% === TABLE ====================================
\begin{table}
\squeezetable
\begin{center}
\begin{tabular}{c|c|c|c|c|c|c|c|c}
%{c|c@{\hspace{3pt}}c@{\hspace{3pt}}c|@{\hspace{5pt}}c@{\hspace{5pt}}c@{\hspace{5pt}}c@{\hspace{5pt}}c@{\hspace{5pt}}c}
\hline
No. & $\langle Q^2\rangle $ & $\langle W\rangle $ &
$\langle \epsilon\rangle$ & $E$ &
P$_{\sss{\mathrm{HMS}}}$ & $\theta_{\sss{\mathrm{HMS}}}$ &
P$_{\sss{\mathrm{SOS}}}$ & $\theta_{\sss{\mathrm{SOS}}}$ \\
 & (GeV/c)$^2$ & GeV & {\small{(for $\Lambda$)}} & GeV & GeV/c &  & GeV/c &  \\
\hline
1 & 0.52 & 1.84 & 0.552 & 2.445 & 0.833 & 29.27$^\circ$ & 1.126 & 13.40$^\circ$
\\
2 & 0.52 & 1.84 & 0.771 & 3.245 & 1.633 & 18.03$^\circ$ & 1.126 & 16.62$^\circ$
\\
3 & 0.52 & 1.84 & 0.865 & 4.045 & 2.433 & 13.20$^\circ$ & 1.126 & 18.34$^\circ$
\\
\hline
4 & 0.75 & 1.84 & 0.462 & 2.445 & 0.725 & 37.95$^\circ$ & 1.188 & 13.42$^\circ$
\\
5 & 0.75 & 1.84 & 0.724 & 3.245 & 1.526 & 22.44$^\circ$ & 1.188 & 17.62$^\circ$
\\
6 & 0.75 & 1.84 & 0.834 & 4.045 & 2.326 & 16.23$^\circ$ & 1.188 & 19.75$^\circ$
\\
\hline
7 & 1.00 & 1.81 & 0.380 & 2.445 & 0.635 & 47.30$^\circ$ & 1.216 & 13.40$^\circ$
\\
8 & 1.00 & 1.81 & 0.678 & 3.245 & 1.435 & 26.80$^\circ$ & 1.216 & 18.20$^\circ$
\\
9 & 1.00 & 1.81 & 0.810 & 4.045 & 2.236 & 19.14$^\circ$ & 1.216 & 20.78$^\circ$
\\
\hline
10 & 2.00 & 1.84 & 0.363 & 3.245 & 0.844 & 50.59$^\circ$ & 1.634 & 13.42$^\circ$
\\
11 & 2.00 & 1.84 & 0.476 & 3.545 & 1.145 & 41.11$^\circ$ & 1.634 & 15.67$^\circ$
\\
12 & 2.00 & 1.84 & 0.613 & 4.045 & 1.645 & 31.83$^\circ$ & 1.634 & 18.14$^\circ$
\\
\hline
\end{tabular}
\caption{Kinematical settings measured in E93-018. Note that there are
three settings of the virtual photon polarization, $\epsilon$, for
each of four values of $Q^2$.  Data were taken in the $\Lambda$ and
$\Sigo$ channels simultaneously.  For ease of discussion, the settings have
been labeled as Point~1 through Point~12 in increasing order of $Q^2$, and
with increasing order of $\epsilon$ within each $Q^2$ setting.
The quantities (P$_{\sss{\mathrm{HMS}}}$,$\theta_{\sss{\mathrm{HMS}}}$)
and (P$_{\sss{\mathrm{SOS}}}$,$\theta_{\sss{\mathrm{SOS}}}$) are the
central momentum and angle settings of the two spectrometers used for
electron and kaon detection, respectively.}
\label{tab:kinematics}
\end{center}
\end{table}
% ==============================================
%
% next file is apparatus.tex

\section{Experimental Apparatus}
\label{sec:apparatus}

\begin{figure}
\begin{center}
\includegraphics[width=3.3in]{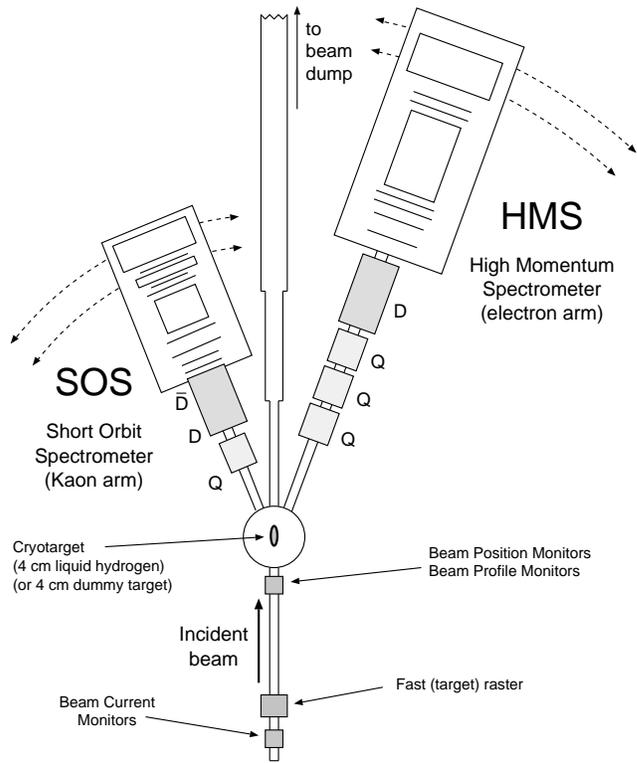}
\caption{Schematic top view of Hall C spectrometer setup showing the
location of the HMS and SOS relative to the target and incident beam.}
\label{fig:hallctop}
\end{center}
\end{figure}

Data collection for experiment E93-018 took place in Hall C
of Jefferson Lab in 1996.  A schematic top view showing the relation
between the spectrometers and the beamline/target chamber is
depicted in Fig.~\ref{fig:hallctop}.  The Continuous Wave (CW,
100\% duty factor) electron beam delivered to Hall C
consisted of 1.67~ps micropulses
spaced approximately 2~ns apart arising from the 499 MHz RF
structure of the accelerator, with beam energies between 2.4 and 4
GeV. The primary target was a (4.36 $\pm$ 0.01)~cm long liquid
hydrogen cryotarget with 0.01~cm aluminum end windows. Background
from the end windows, always less than a few percent, was measured
(and subtracted) using an aluminum target with approximately 
ten times the thickness of the target windows. The integrated beam current
(10-40~$\mu$A) was measured to an accuracy of 1.5\% using a pair
of microwave cavities calibrated with a DC current transformer. In
order to reduce target density fluctuations arising from beam
heating, the beam was rastered in a 2$\times$2~mm$^2$ square
pattern at the entrance to the target.

% === GRAPHICS ====================================
% ==============================================

The scattered electrons were detected in the High Momentum Spectrometer (HMS),
consisting of three superconducting quadrupole
magnets in sequence followed by a superconducting dipole, followed by
a detector package situated near the focal plane of the optical
system. The electroproduced
kaons were detected in the Short Orbit Spectrometer (SOS).  The SOS is a
non-superconducting magnetic spectrometer with one quadrupole (Q) magnet
followed by two dipoles (D and $\overline{\mathrm{D}}$) which share a common
yoke. It was designed with a short flight path in order to allow for
detection of unstable, short-lived particles, such
as kaons or pions, with good efficiency.
Selected properties of the two spectrometers are listed in
Table~\ref{tab:specprop}.

Both spectrometers were equipped with multiwire drift chambers for
particle tracking and segmented scintillator hodoscope arrays for
time-of-flight (TOF) measurement and trigger formation.
Additionally, the HMS had a combination of a gas-filled
threshold \v{C}erenkov detector and a lead-glass calorimeter for
$e$/$\pi^-$ separation, while the SOS had a diffusely-reflecting
aerogel threshold \v{C}erenkov detector ($n=1.034$) for the
purposes of K$^+$/$\pi^+$ separation.  A lucite \v{C}erenkov
detector was also in the detector stack. It was not used in the
trigger or in the present analysis, but was included 
when determining the energy loss
of the kaons passing through the detector. Detection
efficiencies in both spectrometers were dominated by the track
reconstruction efficiency, with additonal small losses due to the
coincidence circuit and the data acquisition dead time (see Table
\ref{tab:corr}).

% === TABLE ====================================
\begin{table}
\squeezetable
\begin{center}
\begin{tabular}{l | c| c}
\hline
 & HMS & SOS \\
\hline
Maximum central momentum & 7.5 GeV/c & 1.75 GeV/c \\
Momentum acceptance & $\pm$10 \% & $\pm$20 \% \\
Ang. Acceptance (in-plane) & $\pm$28 mrad & $\pm$57 mrad \\
Ang. Acceptance (out-of-plane) & $\pm$70 mrad & $\pm$37 mrad \\
Solid Angle (4.36 cm LH$_2$ target) & 6.8 msr & 7.5 msr \\
Optical length & 26.0 m & 7.4 m \\
%\hline\hline
\hline
\end{tabular}
\caption{Selected properties of the HMS and SOS.} \label{tab:specprop}
\end{center}
\end{table}
% ==============================================
% === TABLE ====================================
\begin{table}
\squeezetable
\begin{center}
\begin{tabular}{l|lcc}
\hline
Property & Typical & Random & Scale\\
         & Correction & Error & Error \\
\hline
HMS Tracking efficiency & 0.91 - 0.98  & 0.5 \%  & --- \\
SOS Tracking efficiency & 0.83 - 0.93  & 1.0 \%  & --- \\
HMS Trigger efficiency & 1.0 & 0.1 \%  & --- \\
SOS Trigger efficiency & 1.0 & 0.1 \%  & --- \\
Coincidence efficiency & 0.950 - 0.985  & 0.5 \% & --- \\
TOF $\beta$ cut efficiency & 0.96 - 0.99  & 0.7 \% & --- \\
HMS Elec. Live Time & 0.996 - 1.000 & 0.1 \%  & --- \\
SOS Elec. Live Time & 0.973 - 0.990 & 0.1 \%  & --- \\
Computer Live Time & 0.91 - 1.00 & 0.3 \%  & --- \\
Cointime cut efficiency & 1.0 & 0.1 \% & --- \\
HMS \v{C}erenkov efficiency & 0.998 & 0.2 \% & --- \\
Aerogel cut efficiency & 0.974 & --- & 0.3 \% \\
\hline
HMS Acceptance & --- & --- & 2.0 \% \\
SOS Acceptance & --- & --- & 2.0 \% \\
Cut variation  & --- & 0.7 - 3.1 \% & --- \\
Cross section model & 0.9 - 1.1 & 0.5 \% & 2.0 \% \\
Radiative Correction & --- & 1.0 \%  & 1.0 \% \\
Cut variation ($m_{\sss Y}$) & --- & 0.5 \% & --- \\
Uncert. in $\theta_e$ & & 0.3 - 1.3 \% & \\
Acceptance $\times$ Rad Corr & 1.2 - 1.4 & & \\
\hline
Kaon Absorption & 0.94 - 0.97 & 0.5 \% & 0.5 \% \\
Kaon Decay & 2.5 - 4.0 & 1.0 \%  & 3.0 \% \\
Decay Prod. K$^+$ Mimic& 0.990 - 0.995 & 0.5 \%  & --- \\
\hline
Target Length/Density &  --- & --- & 0.4 \% \\
Target Density Fluct.& 0.992 & 0.4 \% & --- \\
Target Purity & 0.998 & --- & 0.2 \% \\
Charge Measurement & ---  & --- & 1.5 \% \\
%\hline
%Random Subtraction & 2\% to 5\% & 0.5 \% & --- \\
%Endcap Subtraction & 1 - 2\% $\Lambda$                  & 0.5 \%&---\\
%                  & 4 - 8\% $\Sigo$ & \\
\hline \hline
Total & &  2.5 - 4.0 \% &  5.0 \% \\
\hline
\end{tabular}
\end{center}
\caption{Systematic corrections and errors in the E93-018
analysis.} \label{tab:corr}
\end{table}
% ==============================================
%
% next file is dataanalysis.tex

\section{Data Analysis}
%\label{sec:dataanalysis}

The raw data were processed and combined with additional experimental
information such as the momentum and angle settings of the spectrometers,
detector positions, and beam energy to yield particle trajectories,
momenta, velocities, energy deposition, and to perform
particle identification.  Physics variables (such as
$Q^2$, $W$, $\theta_e$, $\mathbf{p_{\sss{K}}}$,
$\theta_{\sss{qK}}$...) were determined for each event at the interaction
vertex, and then yields of electroproduced kaons as a
function of a given subset of these variables were calculated.

\subsection{Kaon Identification}
%\label{sec:kaonPID}

In the SOS spectrometer, the velocities of the detected particles were
calculated using the timing information from the scintillator
hodoscopes. Once the velocities were determined, 
two additional software cuts were implemented to select
kaons out of the proton and pion backgrounds:
a direct cut on the velocity as measured from TOF
information (called $\betatof$), which eliminated
the protons and the majority of the pions, and a cut 
on the number of photoelectrons
detected in the aerogel \v{C}erenkov detector which eliminated 
the remaining pions.
The cut on $\betatof$ was implemented as a cut on the quantity ($\betatof$ $-$
$\betak$), where $\betak$ is the velocity of the detected hadron as determined
from the measured momentum under the assumption that the incident particle
was a kaon, defined as:
\begin{equation}
\betak \equiv
\frac{|{\mathbf{p}}|}{E} = \frac{|{\mathbf{p}}|}
{\sqrt{|{\mathbf{p}}|^2 + m_{\sss{K}}^2}}.
\end{equation}

\begin{figure}
\begin{center}
\includegraphics[width=3.3in]{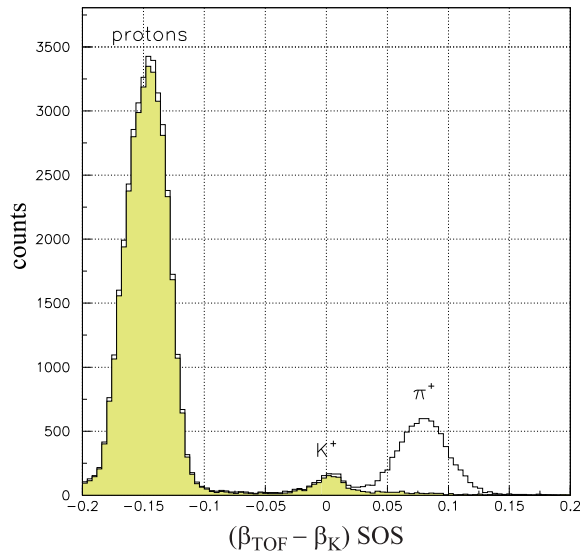}
\caption{A typical $\beta_{\mathrm{tof}} - \beta_{\mathrm{\sss{K}}}$
spectrum for the SOS, shown with (shaded) and without (unshaded)
a software aerogel cut. The data in this figure and 
Figs.~\protect\ref{fig:scointimbeta} and~\protect\ref{fig:scointimcut}
are from kinematic point 2.}
\label{fig:betak}
\end{center}
\end{figure}

The unshaded spectrum in Fig.~\ref{fig:betak} shows the
quantity ($\betatof$ $-$ $\betak$).  In the analysis, a cut was placed at
$|$~($\betatof$ $-$ $\betak$)~$|$~$<$~0.04.

% === GRAPHICS ====================================
% this figure is TOF from 11394... cuts on delta
% kinematic point 2
% and hcer_npe,and no aero/aero=3.5
% ==============================================

The second cut was on the number of photoelectrons detected in the aerogel
\v{C}erenkov.  This eliminated most of the
remaining background pions that the
($\betatof$~$-$~$\betak$) cut did not reject.  The (very few)
pions that survived both of these cuts were eliminated by the
subtraction of the random background (see next section).  The shaded region in
Fig.~\ref{fig:betak} shows what remains after applying a cut requiring less
than 3.5 photoelectrons in the aerogel detector.

\subsection{Coincidence Cuts}
%\label{sec:randombkgd}

\begin{figure}
\begin{center}
\includegraphics[width=3.3in]{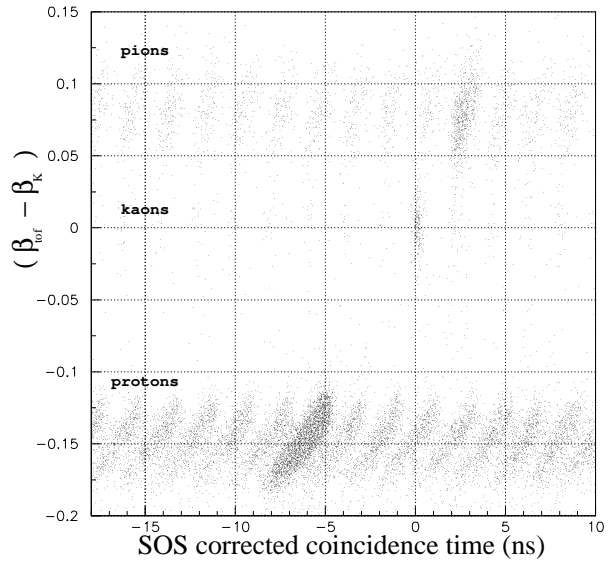}
\caption{Plot
of $(\beta_{\mathrm{tof}} - \beta_{\mathrm{\sss{K}}})$ vs.~SOS
corrected coincidence time. Visible are three horizontal bands
corresponding to protons, kaons, and pions. The coincidence time
offset is chosen such that the in-time kaons appear at 0 ns. The
correlation between coincidence time and 
$(\beta_{\mathrm{tof}} - \beta_{\mathrm{\sss{K}}})$ in the proton
bands reflects the range of proton velocities that could create a
random coincidence with the electron arm within the trigger timing 
window.}
\label{fig:scointimbeta}
\end{center}
\end{figure}

The relative timing between the HMS and SOS signals was used to
identify true kaon-electron coincidences. This coincidence time
was corrected to account for variations in flight time arising
from variations in particle velocity and path length through the
spectrometers. An arbitrary offset was added such that events in
which the electron and kaon originated from the same beam bunch
would have a time of 0 ns. Figure~\ref{fig:scointimbeta} shows
($\betatof$ $-$ $\betak$) for the SOS plotted versus the corrected
coincidence time for a single run, without having applied the 
previously described kaon
identification cuts. Three horizontal bands of electron
coincidences with protons, kaons, and pions are clearly
identified, with the in-time pion and proton peaks offset from the
in-time kaon peak by at least one beam bunch.  Random
coincidences, resulting from an electron and hadron from different
beam bunches, have a coincidence time that is offset from the
in-time peak by a multiple of $\sim$2 ns. 

% === GRAPHICS ====================================
% this figure is from 11394... cuts on delta and hcer_npe and all cuts
% ==============================================

% === GRAPHICS ====================================
% this figure is from 11394... cuts on delta and hcer_npe and all cuts
\begin{figure}
\begin{center}
\includegraphics[width=3.3in]{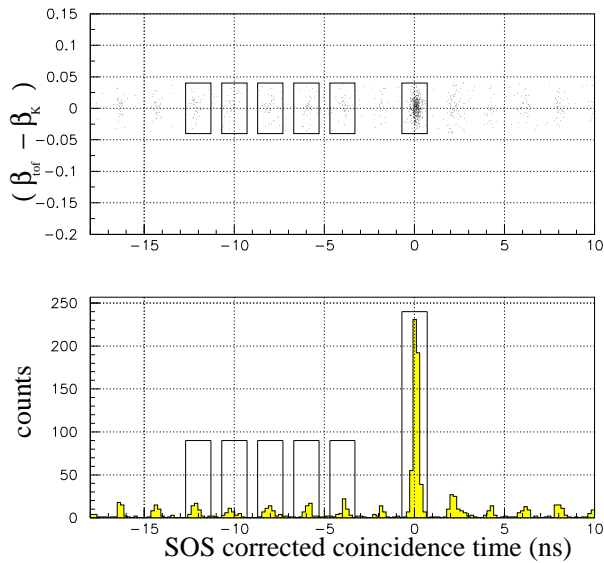}
\caption{The upper panel
shows $(\beta_{\mathrm{tof}}-\beta_{\mathrm{\sss{K}}})$ vs. SOS
corrected coincidence time, after applying cuts on
$(\beta_{\mathrm{tof}}-\beta_{\mathrm{\sss{K}}})$ and the aerogel.
The single boxed region to the right is the in-time peak, and the
five boxed regions to the left contain random coincidences.  The
lower panel shows the one-dimensional spectrum of SOS corrected
coincidence times corresponding to the upper panel.}
\label{fig:scointimcut}
\end{center}
\end{figure}
% ==============================================

After applying the ($\betatof$ $-$ $\betak$) and aerogel cuts, the
distribution shown in Fig.~\ref{fig:scointimbeta} reduces to
that shown in Fig.~\ref{fig:scointimcut}(a). In
Fig.~\ref{fig:scointimcut}(b) the one-dimensional projection of
the upper half is shown. The in-time peak dominates over the
purely random coincidences.  The data were cut on the coincidence
time around the peak, at $ | \, \mathrm{cointime(SOS)} \, | < 0.65
\, \, \mathrm{ns}$. The random background was estimated by
averaging over five bunches to the left of the in-time peak, and
then subtracting it from the in-time yield to give the true kaon
electroproduction yield.

\subsection{Missing Mass Reconstruction}

Once true $e$-$K$ coincidences were identified,
the missing mass, $m_{\sss{Y}}$, of the
produced hyperon was reconstructed from measured quantities as
\begin{equation}
\label{eq:mmass}
m_{\sss Y}^{2} =
  -Q^{2} + m_{\sss K}^{2} +
  2 \, \nu \, m_{p} - 2 \, E_{\sss K} ( \nu + m_{p} ) -
  2 \, \mathbf{q} \cdot \mathbf{p_{\sss K}} \, \, ,
\end{equation}
where $(E_{\sss K},\mathbf{p_{\sss K}})$ is the kaon four-vector, and
$(\nu,\mathbf{q})$ is the virtual photon four-vector.

% === GRAPHICS ====================================
\begin{figure}
\begin{center}
\includegraphics[width=3.3in]{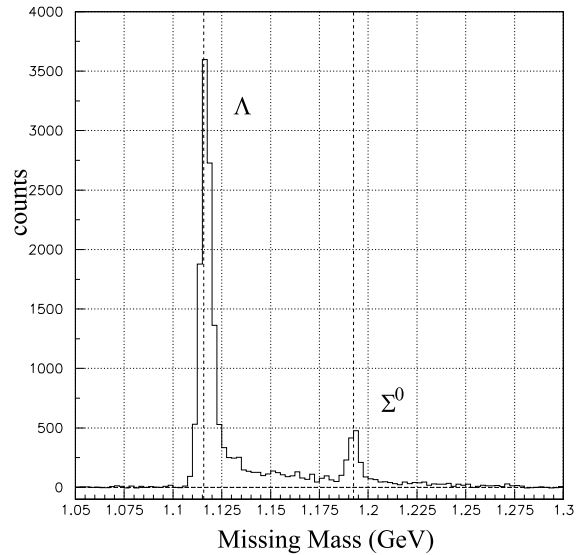}
\caption{An example of a missing mass spectrum for $\peek$Y showing the
$\Lambda$ and $\Sigo$ peaks and radiative tails.  The vertical dashed
lines are
located at the accepted $\Lambda$ and $\Sigo$ masses. Note that this spectrum
has already been corrected for random and target endcap yields.}
\label{fig:mmass}
\end{center}
\end{figure}
% ==============================================

Figure~\ref{fig:mmass} is a histogram of the calculated missing mass,
showing peaks at the $\Lambda$ and $\Sigo$ masses. The tail sloping off
toward higher missing mass from each peak is due to the effects of
radiative processes.
A cut of $(1.100 < m_{\sss{Y}} < 1.155)$~GeV was
used to identify events with a $\Lambda$ in the final state, and
$(1.180 < m_{\sss{Y}} < 1.230)$ was used
for the $\Sigo$ final state. The fraction of events lost due to
the cut were accounted for in the data/Monte Carlo ratio used
to determine the cross section. The $\Sigo$
analysis also required subtraction of the $\Lambda$ events 
in the radiative tail under the $\Sigo$ peak.
After application of all cuts and identification of true $\peek$Y events, the
measured yields were corrected for losses due to detection efficiencies, 
kaon decay, and kaon absorption through the target and
spectrometer materials.  A summary of these corrections and their associated
errors can be found in Table \ref{tab:corr}.

\label{sec:kdecaycorr}
One of the largest corrections arose from the fact that kaons are unstable and
have a short mean lifetime. As a result, a large fraction of the kaons
created at the target decay into secondary particles before they
can be detected. The survival fraction of the detected kaons is
\begin{equation}
\label{eq:expdecay}
\frac{N_{\mathrm{detected}}}{N_{\mathrm{at\, target}}}
= \exp{{\left(-\frac{m_{\sss K}\, d}{p_{\sss K}\, \tau}\right)}}\, ,
\end{equation}
where $d$ is the distance traveled to the detector and $\tau
\approx 12.4$~ns ($c\tau \approx 3.713$~m) is the
mean lifetime of a kaon at rest~\cite{PDB}.  
The survival fraction varied between 0.25 and 0.4 for the range
of kaon momenta detected here.  Although the mean kaon decay
correction is large, its uncertainty is small because both the
kaon's trajectory length and momentum were accurately determined
event by event by the SOS tracking algorithm. 
An additional few percent correction was
applied to account for the possibility that the decay products
might actually be detected, mimicking an otherwise lost kaon
event. This correction was estimated through the use of a Monte
Carlo simulation of the six most likely decay processes as listed
in~\cite{PDB}. Further details regarding the data
analysis can be found in ~\cite{rmmthesis}.

\section{Monte Carlo Simulation}
\label{sec:mc}

In order to extract cross section information from the
data, a detailed Monte Carlo simulation of the experiment was
developed. The code was 
largely based on that developed in Ref.~\cite{makinsthesis}, which used the
Plane Wave Impulse Approximation (PWIA) to model $A(e,e'p)$ for various
nuclei. The simulation included radiation, multiple scattering, 
and energy loss from passage through materials. It was updated with
optical models of the HMS and SOS spectrometers and compared in detail
with Hall C $(e,e')$ and $(e,e'p)$ data. For the present analysis it
was additionally modified to simulate kaon electroproduction as well
as kaon decay in flight. In the discussions that follow, the term
``data'' always refers to the measured experimental data yields, 
and the term ``MC'' always refers to the simulated events and yields.

\subsection{Event Generation}

In the MC event generator, a random target interaction point
is selected, consistent with the target length and raster amplitudes.
The beam energy is then chosen about a central (input) value with
a resolution $dE/E = \pm$0.05\%. 
Events are randomly generated in the phase space including the
spectrometer angles and electron momentum, from which the laboratory
quantites $p_e$, $\theta_e$, $\phi_e$, $\theta_{\sss K}$
and $\phi$ are computed. From the distribution of events, the
phase space factor $\Delta V^5_{\mathrm{gen}} \equiv 
\Delta E' \Delta \Omega'_e \Delta \Omega_{\sss K}$ is also computed.
Each event is generated with unit cross section, with limits 
%The laboratory quantities $p_e$, $\theta_e$, $\phi_e$, $\theta_{\sss K}$
%and $\phi$ are randomly generated for each event, with unit cross
%section% in the phase space $\Delta V^5_{\mathrm{gen}} \equiv 
%\Delta E' \Delta \Omega'_e \Delta \Omega_{\sss K}$ and with limits
that extend well beyond the physical acceptances of the spectrometers
even after the effects of energy loss, radiation and multiple scattering.
From the scattered electron laboratory variables, the invariant
quantities $Q^2$ and $W$ can be completely specified,
as well as the kaon CM production angle $\theta_{\sss qK}^{*}$, 
and the virtual photon polarization $\epsilon$.

For the kaon side, the hyperon $Y$~=~$(\Lambda ,\Sigo)$ being 
generated is specified in the Monte Carlo input file.  Because the
kaon-hyperon production is a two-body reaction, the momentum of the outgoing
kaon is fixed by knowledge of $W$. In the CM frame this relationship
is 
\begin{equation} 
\label{eq:pkcm}
|{\mathbf{p^{*}_{\sss K}}}| = \sqrt{ \left(
\frac{(W^{2} + m_{\sss K}^{2} - m_{\sss Y}^{2})}{2 \, W}\right)^{2}
- \, m_{\sss K}^{2} } \, \, .
\end{equation}
The remaining components of the kaon four-vector 
can be determined using the kaon CM angles
$(\theta_{\sss qK}^{*}, \phi)$. In the Monte Carlo code, the
laboratory kaon momentum is instead computed, from which CM quantities
at the interaction vertex are calculated. 
The Jacobian that relates the kaon CM solid angle 
$\Delta \Omega_{\sss K}^{*}$ to its laboratory counterpart
is also specified since the reported cross sections are in the
CM frame.

\label{sec:radcor}
The radiative correction routines supplied in the 
original ($e$,$e'p$) code
of Ref.~\cite{makinsthesis} were based on
the work of Mo and Tsai \cite{motsai}, extended to be valid for a
coincidence framework.  Details of the corrections are
documented in~\cite{ent2002}. The
procedure outlined in this reference was implemented for quasielastic
proton knockout. In the present work, the same procedure was used with two
modifications to consider kaon production: the mass of struck and
outgoing particle was changed to $m_K$. This is appropriate for the 
pole contribution to the radiation, and better reproduces the data.
In addition, the vertex reactions were modified to follow a 
parameterization of the ($e$,$e'K$) cross section (discussed below).

% === GRAPHICS ====================================
% taken from ~/nt/022899/pt9_sc.hbook
\begin{figure}
\begin{center}
\includegraphics[width=3.3in]{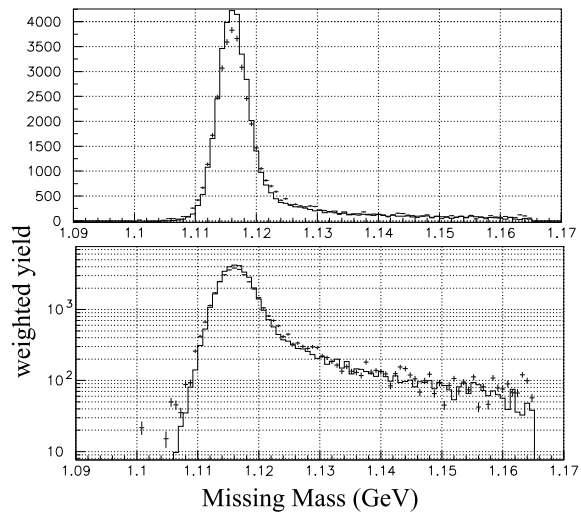}
\caption{Yield spectrum of missing mass comparing MC (solid line) to data
(crosses) for $p$($e$,$e'K^+$)$\Lambda$, on both linear (top panel) and 
logarithmic
(bottom panel) scales, normalized by the extracted cross section.
The data are from kinematic point 9.}
\label{fig:mmassdatamc}
\end{center}
\end{figure}
% ==============================================

Figure~\ref{fig:mmassdatamc} shows the cross section weighted MC
calculation of its yield, including the radiative 
tail (solid line), plotted on top of
the measured data (crosses), on both linear and logarithmic
scales.  Although the resolution of the MC peak is slightly
narrower than that of the data (a result of the spectrometer
model), overall the agreement between the MC and data is quite
good. The missing mass cut used to define the $\Lambda$ (1.100 $<
m_{\sss Y} <$ 1.155 GeV) was sufficiently wide that the extracted
cross section was insensitive at the level of 0.5\% to variations
in the cut.

\begin{figure}
\begin{center}
\includegraphics[width=3.3in]{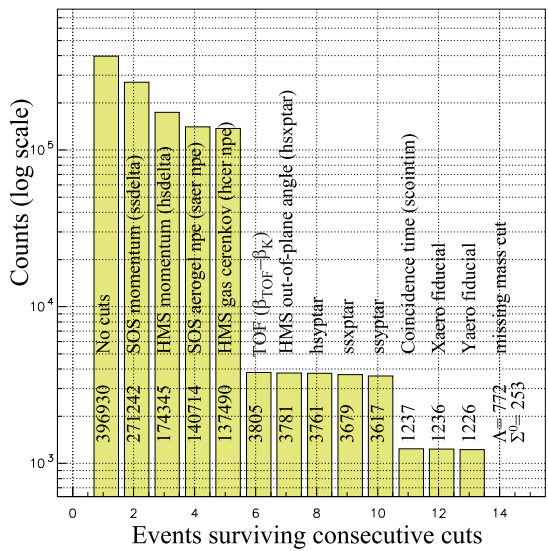}
\caption{Histogram showing the cumulative effects of applying the cuts (note
logarithmic scale).  Shown are the number of counts surviving after
applying the cuts as indicated from left to right, resulting in a final yield
of 772 $\Lambda$ and 253 $\Sigo$. The data are from one run of kinematics
3.}
\label{fig:cuteff}
\end{center}
\end{figure}

Figure~\ref{fig:cuteff} shows the sequential effect of placing the cuts on the
experimental data from a single run. In the analysis, all of the cuts were
systematically varied and their effects on the cross section noted.  The
standard deviation of the resulting cross sections was used as an estimate of
the systematic error on the quoted cross section due to each cut.
With the exception of a cut on $\thcm$, the cut values used were not
kinematic dependent and are listed in table~\ref{tab:cuts}.

% === TABLE ====================================
\begin{table}
\squeezetable
\begin{center}
\begin{tabular}{l|lcc}
\hline
SOS momentum  & $\vert\Delta P_K/P_k \vert <$ 17\% \\
HMS momentum & $\vert\Delta P_e/P_e \vert <$ 8\%  \\
HMS out of plane angle & $\vert dx/dz\vert_{HMS} <$ 0.08 rad \\
HMS in-plane angle & $\vert dy/dz\vert_{HMS} < $ 0.04 rad \\
SOS  out of plane angle & $\vert dx/dz\vert_{SOS}< $ 0.04 rad \\
SOS in-plane angle & $\vert dy/dz\vert_{SOS} < $ 0.07 rad  \\
aerogel X (dispersion dir.) & -49.0 cm $<$ X $<$ 46.0 cm  \\
aerogel Y & -14.0 cm $<$ Y $<$ 18.0 cm & \\
Missing mass cut, $\Lambda$ only & 1.100 GeV $<$ $m_{\sss Y}$ $<$ 1.155 GeV \\
Missing mass cut, $\Sigma_0$ only& 1.180 GeV $<$ $m_{\sss Y}$ $<$ 1.230 GeV \\
\hline
$\thcm$ cut & \\
\hline
kin 1/2/3 & 10/11/12$^\circ$ \\
kin 4/5/6 & 10/12/14$^\circ$ \\
kin 7/8/9 & 10/12/14$^\circ$ \\
kin 10/11/12 & 12/14/16$^\circ$ \\
\hline\hline
\end{tabular}
\end{center}
\caption{Fiducial cuts placed on both the data and Monte Carlo 
in determining the data/MC ratio. Additional cuts, as described in
the text, were placed on data quantities, such as the number of aerogel
and gas \v{C}erenkov photoelectrons, coincidence time, and kaon velocity.} 
\label{tab:cuts}
\end{table}
% ==============================================

% === GRAPHICS ====================================
% ==============================================

\subsection{Equivalent Monte Carlo Yield}

The equivalent experimental yield given by the MC can be expressed as
\begin{eqnarray}
Y_{\mathrm{MC}} = L_{\sss H} &\times & \int
\left[\frac{d^5\sigma}{dE' d\Omega'_e d\Omega_{\sss K}^{*}}\right]
\left(\frac{d\Omega_{\sss K}^{*}}{d\Omega_{\sss K}}\right)
\nonumber \\
&\times &
A(d^{\,5}V) \, R(d^{\,5}V) \, dE' d\Omega'_e d\Omega_{\sss K} \, ,
\end{eqnarray}
where $L_{\sss H}$ is the experimental luminosity, $A$ is the
acceptance function of the
coincidence spectrometer setup, and $R$ is the radiative correction.  The
experimental luminosity is given by
\begin{equation}
L_{\sss H} = C_{\mathrm{eff}} \, N_{\mathrm{beam}} \, N_{\mathrm{tgt}} \, , 
\end{equation}
where $C_{\mathrm{eff}}$ is a multiplicative factor containing all the global
experimental efficiencies (such as the tracking efficiency, dead times, etc.),
and $N_{\mathrm{beam}}$
and $N_{\mathrm{tgt}}$ are the number of incident electrons and the number
of target nucleons/cm$^2$, respectively.

Substituting the virtual photoproduction cross section 
(Eq.~\ref{eq:virtphot}) for the full electroproduction 
relation results in
\begin{eqnarray}
\label{eq:virtxsect}
Y_{\mathrm{MC}} = L_{\sss H} &\times & \int 
\biggl[ \Gamma_0(E^\prime, \Omega^\prime_e) \,
\left(\frac{d^{\,2}\sigma}{d\Omega_{\sss K}^{*}}\right) 
\left(\frac{d\Omega_{\sss K}^{*}}{d\Omega_{\sss K}}\right)
\biggr] \nonumber \\
&\times & A(d^{\,5}V) \, R(d^{\,5}V) \,
dE^\prime d\Omega^\prime_e d\Omega_{\sss K}.
\end{eqnarray}

In order to eventually extract cross sections by comparison with
data, the MC yield was weighted by a cross section model that combines
a single global factor $\sigma_0$, representing the cross section at 
specified values of $Q^2$, $W$ (denoted ($Q_0^2$,$W_0$))
and $\avgthcm=0^\circ$,
with a function representing the event-by-event variation of
the cross section across the experimental acceptance. The
$\thcm$ behavior was represented by a variation in $t$, with
$\avgthcm=0^\circ$ corresponding to the minimum accessible value
of $t$, denoted $t_{min}$. 
The cross section model was based on previous data. 
This procedure is equivalent to replacing the cross section
in Eq.~\ref{eq:virtxsect} with the factorized expression
\begin{eqnarray}
\label{eq:scalingfndef0} 
\left(\frac{d^{\,2}\sigma}{d\Omega_{\sss K}^{*}}\right)  
\equiv \sigma_0 \times\frac{f_Q(Q^2) f_W(W) f_t(t)}
{f_Q(Q_0^2) f_W(W_0) f_t(t_{min})}\, 
\end{eqnarray}
where the various functions will be described below. 
The data constrain the choice of cross section models to those
with relatively little variation 
in $Q^2$, $W$, or $t$ across the acceptance. As a result,
the extracted cross section is not very sensitive to detailed
behavior of the model.

The integral in Eq.~\ref{eq:virtxsect} 
is evaluated numerically via the Monte Carlo simulation, with
each Monte Carlo event being appropriately weighted with the radiative
corrections, virtual photon flux, and the ($Q^2$,$W$,$t$) dependent
terms of the cross section model. The
influence of the acceptance function, $A$, arises through the fraction of
generated events that successfully traverse the spectrometers and
are reconstructed. 

The MC equivalent yield then reduces to
\begin{eqnarray}
\label{eq:extract}
Y_{\mathrm{MC}} &=& L_{\sss H} \times\sigma_0
\times \Delta V^5_{\mathrm{gen}} \nonumber \\
&\times &
\left(\frac{(\text{Number of MC successes})_{\mathrm{weighted}}}
{\text{Number of MC tries}}\right)\, , 
\end{eqnarray}
\label{sec:xsect-extract} 
and by adjustment of $\sigma_0$ 
such that the yields of MC and measured data are equal 
(i.e., $Y_{\mathrm{MC}} = Y_{\mathrm{DATA}}$), 
the cross section $\sigma_0$ at the specified kinematics
is determined.

\subsection{Cross Section Model}
%\label{sec:scaling}

The previously existing world data were used to account for the cross
section behavior across the acceptance.  In 
Bebek~\textit{et al.}~\cite{bebek-lt}, where cross sections 
at $\thcm$=0$^\circ$ were presented, 
the behavior of the cross section was parameterized as:
\begin{equation}
\label{eq:bebekqw}
\frac{d^{\,2}\sigma}{d\Omega_{\sss K}^{*}} \sim f_Q(Q^2)
\times f_W(W) \, ,
\end{equation}
where
\begin{equation}
\label{eq:bebekq}
f_Q(Q^2) = \frac{1}{(Q^{2} + X)^{2}}
\end{equation}
with $X = 2.67$ for the $\Lambda$ channel, and $X = 0.79$ for the
$\Sigo$, and
\begin{equation}
\label{eq:bebekw}
f_W(W) = \frac{|\mathbf{p^{*}_{\sss K}}|}{W ( W^{2} -
m_{p}^2)} .
\end{equation}

In a recent analysis of another set of kaon electroproduction
data~\cite{DMKthesis},
it was shown that data at values of $W$ near the production
threshold for the $\Lambda$ cross section were not accurately reproduced by
the $W$ dependence in Eq.~\ref{eq:bebekw}. In that analysis, a function of
the following form was proposed, motivated by
the hypothesis that there are possible resonance contributions to
the cross section at lower $W$:
\begin{equation}
\label{eq:dmkw}
f_{\rm res}(W) = C_1 f_W(W)  +
C_2 \frac{\, \, A^2 \, B^2}{(W^2 - A^2)^2 + A^2 B^2}
\end{equation}
with  $A$=$1.72$~GeV, $B$=$0.10$~GeV,
$C_1$=$4023.9$~GeV$^2$~nb/sr, and $C_2$=$180.0$~GeV$^2$~nb/sr.
This modified function was used here for the $\Lambda$ channel. 
In the $\Sigma$ channel, no such modification was found to be 
necessary to fit the existing data, so the original Bebek 
parameterization was used.

The $\thcm$ behavior was estimated using the results of
Brauel,~\textit{et al.}~\cite{brauel}:
\begin{equation}
2\pi \frac{d^{\,2}\sigma}{dt d\phi}
\sim
e^{- \xi |t|} \, ,
\label{eq:tdep}
\end{equation}
where $\xi = 2.1$ for the $\Lambda$, and $\xi = 1.0$ for the $\Sigo$,
with the Mandelstam variable $t$ defined as
\begin{equation}
\label{eq:t}
t = -Q^{2} + m_{\sss K}^{2}
     - 2 \, E_{\sss K}^{*} \nu^{*} + 2 \, |\mathbf{q^{*}}|
     |\mathbf{p_{\sss K}^{*}}| \cos \thcm \, \, ,
\end{equation}
where $(E_{\sss K}^{*},\mathbf{p_{\sss K}^{*}})$ is the kaon four-vector in the
CM frame, and $(\nu^{*},\mathbf{q^{*}})$ is the virtual photon four-vector in
the CM frame. Using Eq.~\ref{eq:t} one can relate
the $t$ behavior to the $\thcm$ behavior as
\begin{equation}
f(\thcm) = \frac{1}{2\pi} \, e^{- \xi |t|} \left(2 |\mathbf{q^{*}}|
|\mathbf{p_{\sss K}^{*}}|\right) \, \, .
\label{eq:thetascale}
\end{equation}
At $\avgthcm = 0^{^{\circ}}$, the variable $t$ becomes $t_{min}$ given by
\begin{equation}
\label{eq:tmin}
t_{min} = -Q^{2} + m_{\sss K}^{2}
- 2 \, E_{\sss K}^{*} \nu^{*} + 2 \, |\mathbf{q^{*}}| |\mathbf{p_{\sss K}^{*}}|
\, \, ,
\end{equation}
resulting in a functional form of
\begin{equation}
\label{eq:tmin2}
\frac{f_t(t)}{f_t(t_{min})} 
=\frac{f(\thcm)}{f(\thcm=0^{^\circ})} =
%\frac{e^{-\xi|t|}}{e^{-\xi|t_{min}|}} = 
e^{-\xi(|t| - |t_{min}|)}
= f_t(t-t_{min})\, .
\end{equation}
Note that $t_{min}$ is a function of $Q^{2}$ and $W$
through its dependence on $\nu^*$.

% === GRAPHICS ====================================
\begin{figure}
\begin{center}
\includegraphics[width=3.3in]{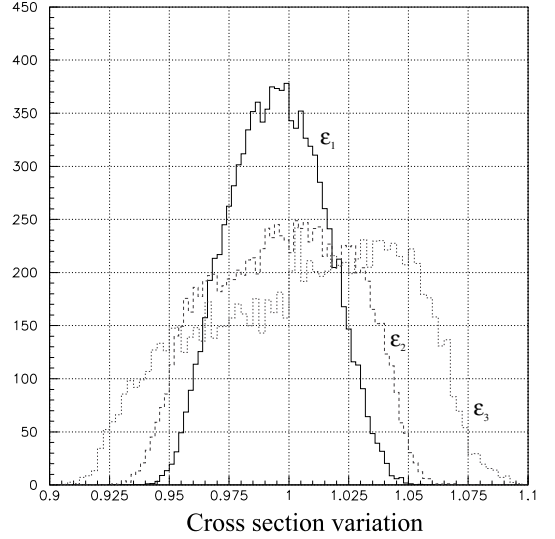}
\caption{The cross section variation compared to its central value
for the three values of $\epsilon$ at $Q^2 =
0.52$~(GeV/c)$^2$ (Points 1, 2, and 3).}
\label{fig:scalingfactor}
\end{center}
\end{figure}
% ==============================================

In order to extract cross sections at the specific value of 
$\avgthcm = 0^{^{\circ}}$, each event in the MC was weighted 
by the $Q^2$ and $W$ functions (defined in Eqs.~\ref{eq:bebekq}
- \ref{eq:dmkw}), and by the $t$-dependent 
function in Eq.~\ref{eq:tmin2}. Cross sections are thus presented
not at fixed values of $t$ but instead at $t_{min}(Q^2, W)$. 

Figure~\ref{fig:scalingfactor} shows a histogram of the
variation of the cross section weighting factor 
about its central value, on an event-by-event basis
corresponding to the three values of $\epsilon$ at the
$Q^2$~=~0.52~(GeV/c)$^2$ setting.  The variation 
is largest at this lowest $Q^2$, and is essentially
always less than $\pm$8\%. The dependence of the extracted cross
section on various deviations to this model was investigated in 
depth and the maximum observed effect was 2.3\%.

\subsection{Differences from the Previous E93-018 Analysis}

As discussed in the introduction, the cross sections for the $\Lambda$
channel that are presented here are significantly different than the
earlier published data of Ref.~\cite{gnprl}. There are four main 
differences between the two analyses, three of which are somewhat
trivial in nature. The first two have little influence on the 
unpolarized cross sections but are epsilon-dependent and therefore
do affect the L/T separated data. Firstly, the radiative correction 
factors applied in Ref.~\cite{gnprl} 
were assumed to be constant for each kinematic
setting at fixed $Q^2$, whereas they in fact vary by about 5\%
with $\epsilon$.  Secondly, in Ref.~\cite{gnprl} it was 
assumed that the contributions from
the interference structure functions $\sigma_{LT}$ and $\sigma_{TT}$
cancel within the acceptance and so no cut in $\theta^*_{qK}$ was 
applied. Since the acceptance is $\epsilon$ dependent, this 
assumption introduces an $\epsilon$-dependent bias. (With no
$\thcm$ cut, the $\phi$ acceptance is not uniform, and
the interference terms will give a net contribution to the cross 
section.) In the analysis presented here, such a cut was applied, and for
each kinematic setting it was chosen as a compromise between optimizing 
the uniformity of the $\phi$ coverage and minimizing 
the effect of the cut on the extracted 
cross sections. Since the $\phi$-dependence of the acceptance 
arises naturally in the MC, the $\phi$-dependence of the acceptance
was mostly removed in the data/MC ratio, and any residual
dependence was attributed to the interference terms in
the cross section.

Finally, in Ref.~\cite{gnprl}, a jacobian 
to convert kaon momentum to missing mass in the laboratory system was 
erroneously applied. The value of this jacobian varied with $Q^2$
from 1.95 to 2.8, but its influence was largely cancelled by
the fourth difference which comes from how the phase-space acceptance
for the detected kaons was handled. This last contribution could 
have a bearing on comparisons of the data reported here
with calculations and with other kaon electroproduction
experiments. 

In principle, in a simulation of the
$p(e,e'K^+)\Lambda$ reaction with the missing mass held fixed, 
and at fixed kaon production angle in the laboratory,
there are two solutions to Eq.~\ref{eq:pkcm}, corresponding to 
forward and backward going kaons in the CM frame. 
The analysis of Ref.~\cite{gnprl} made the fundamentally correct 
assumption that either of the two found solutions is possible. 
This is true in general, and is an appropriate assumption for
a kaon electroproduction experiment in a large acceptance device. 
However, it is inconsistent with the generally biased preference 
to detect the ``forward''-going kaon due to the limited momentum
acceptance magnetic spectrometer. The procedure used in Ref.~\cite{gnprl} 
to account for both kaon momentum solutions led to an increase in the 
assumed phase space by a factor of two, and thus a reduction in the 
cross section of the same amount. This factor largely cancelled the 
effect from the jacobian.

One might argue that an experiment carried out with 
limited acceptance detectors does not truly 
measure an exclusive five-fold laboratory differential cross section, 
but rather a cross section solely related to the forward-going kaons
in the center-of-mass frame. If one had perfect knowledge of the 
kaon electroproduction process,  a simulated 
experiment taking the ``backward''-going kaons into account could
be carried out, from which one could obtain experimental laboratory
cross sections that can be directly
compared with theoretical calculations. Alternatively, additional
experimental configurations could be chosen to measure the 
``backward''-going electroproduction cross sections. 
Obviously, large acceptance devices
do not encounter this problem and have an advantage here. 
However, even in a complete experiment, it would be
necessary to separate out the forward and backward going kaons,
which correspond to different CM angles, when 
converting from the measured lab cross sections to the desired CM values.

In the present analysis, backward-going kaons were simulated 
and found to be well outside of the momentum acceptance of the 
kaon spectrometer, therefore taking only the forward-going 
solution was in fact the consistent way
to match the true experimental conditions. It should also be noted that all 
previous electroproduction experiments with a magnetic spectrometer setup
have reported exclusive five-fold differential electroproduction cross sections
with the same biased preference. 

%=====================================================
% next file is results.tex

\section{Results}

The data and MC were binned in $\phi$ in order to study the 
effect of potential contributions from the interference terms
to the extracted cross sections prior to carrying out the
L/T separation.  Cross sections were extracted by forcing the 
ratio of data to MC yields to be unity through adjustment of the overall
normalization factor $\sigma_0$ in the MC bin-by-bin. After a first pass
through the analysis, both the data and MC yields were stored
in 8 bins of $\phi$. The ratio of data/MC was
calculated in each bin yielding a zeroth order cross section for
that bin. The procedure was then iterated, applying the extracted
$(n-1)^{th}$ order cross section as a weighting factor for the
yields in each $\phi$ bin using the generated values of the MC
$\phi$ as the bin index. Typically the extracted cross section
stabilized to within 0.1\% of its value in three iterations. These
final bin-by-bin values were fitted to a constant plus a harmonic
$\phi$ dependent function of the form $A + B\cos\phi + C\cos2\phi$
in order to extract $A = \sigma_{\sss{T}} +
\epsilon\sigma_{\sss{L}}$. Note that with a single $\phi$ bin, the $\phi$
dependence should naturally cancel if the $\phi$ acceptance is
uniform. This was true provided that the accepted events are restricted 
to forward values of $\theta_{\sss qK}^{*}$. 
The extracted cross sections from a single bin in $\phi$ compared
with 8 bins were unchanged at the level of 0.5\% for the $\Lambda$
channel and at the level of 1.5\% for the $\Sigma_0$ channel.
  
The choice of cut in $\theta_{\sss qK}^{*}$ was kinematic dependent,
and the values used are in Table~\ref{tab:cuts}. 
The $\phi$-dependent terms could not
be quantitatively extracted due to the low statistics per bin and
the poor $\phi$ reconstruction 
resolution for these small values of $\theta_{\sss qK}^{*}$.
However, the amplitude of $\cos\phi$ ($\cos 2\phi$) term was
typically 10\% (5\%) of the unpolarized cross section.

\begin{figure}
\begin{center}
\includegraphics[width=3.3in]{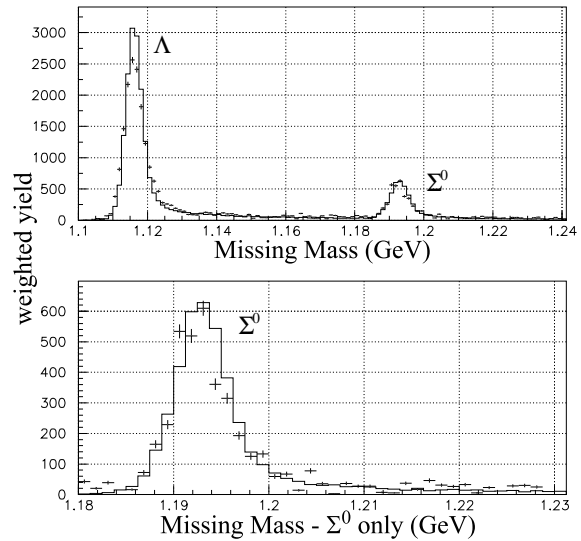}
\caption{Use of the MC to correct for the $\Lambda$ radiative tail
below the $\Sigo$ missing mass peak.
The upper panel shows the combination of the
$\Lambda$-specific and $\Sigo$-specific MC simulations (solid histogram)
plotted on top of the data missing mass (crosses).
The lower panel shows the remaining $\Sigo$ data after
subtracting the $\Lambda$-specific MC, with the $\Sigo$-specific MC
superimposed. The MC is normalized using the extracted $\Lambda$ and
$\Sigo$ cross sections. The data are from kinematic point 3.}
\label{fig:lambkgd}
\end{center}
\end{figure}

The procedure for extracting the $\Sigo$ cross section was similar to that for
the $\Lambda$, except that the $\Sigo$ yield was also corrected for the
$\Lambda$ radiative tail beneath the $\Sigo$ peak in the missing mass spectrum.
The $\Lambda$-specific MC was used to determine the number of background
$\Lambda$ counts that were within $\Sigo$ cuts. The $\Lambda$-specific MC was
weighted with the extracted $\Lambda$ cross section, binned in the same manner
as the data, and was subtracted from each data bin.  The upper half of 
Fig.~\ref{fig:lambkgd} shows the combination of the $\Lambda$-specific and
$\Sigo$-specific MC simulations plotted on top of the data missing mass.
The lower half of the figure shows the remaining $\Sigo$ data after subtracting
the $\Lambda$-specific MC, with the $\Sigo$-specific MC superimposed.
Varying the $\Lambda$ cross section in the $\Sigo$ extraction analysis
by $\pm$10\% resulted in changes of less than 2\% in
the $\Sigo$ cross section.  While the $\Lambda$ cross sections were 
typically determined to better than $\pm$10\%, the contamination of
the $\Lambda$ events in the $\Sigo$ yield comes from events which have
undergone significant radiation and therefore the $\Lambda$ yield in
that region is likely more sensitive to the details of the model
in the Monte Carlo. An additional scale uncertainty proportional to the
$\Lambda$ cross section uncertainty was thus applied to the $\Sigo$ results.

% === GRAPHICS ====================================
% ==============================================

Typical values for all corrections to the data and/or the MC,
along with the resulting systematic errors in the cross section,
are shown in Table~\ref{tab:corr}.  The statistical errors for the
various settings ranged from 1.0$-$3.1\% for the $\Lambda$, and
from 4.8$-$15\% for the $\Sigo$. The systematic errors are broken
down into ``random'' and ``scale'' errors. Since random errors
affect each kinematic setting in an independent manner, they were
retained for the linear fit of the L/T separation while the global
errors that result in an overall multiplicative factor to the data
were ignored for the fitting procedure, and applied as a scale uncertainty
in the individual L/T cross sections.  The sources of the errors
in Table~\ref{tab:corr} are discussed in detail in
\cite{rmmthesis}.

\subsection{$\Lambda$ and $\Sigo$ cross sections}
\label{sec:lamresults}

\begin{figure} 
\begin{center}
\includegraphics[width=3.3in]{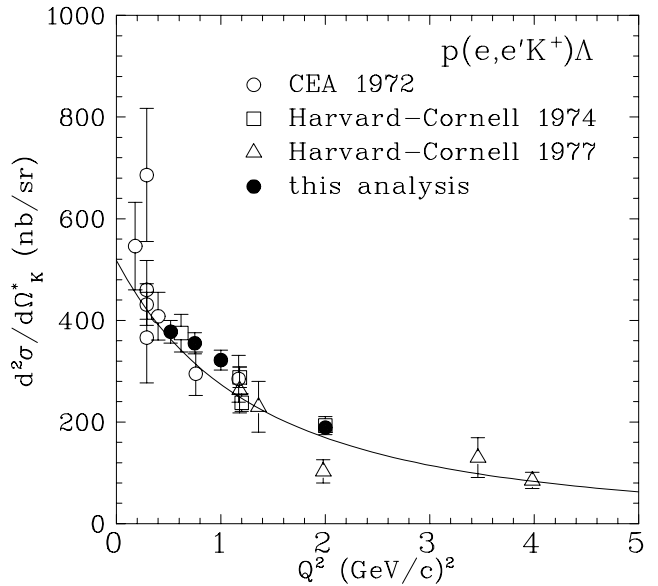}
\caption{Previous world data (open symbols: 
circles~\protect\cite{brownworlddata}, squares~\protect\cite{bebek-74},
diamonds~\protect\cite{bebek-prd0277}) with the addition of the
highest-$\epsilon$ results of this analysis (solid points) scaled to 
$W$~=~2.15~GeV, $\avgthcm=0^\circ$, for the $p(e,e'K^+)\Lambda$
unseparated cross sections.}
\label{fig:world-lam-plus}
\end{center}
\end{figure}

\begin{figure} 
\begin{center}
\includegraphics[width=3.3in]{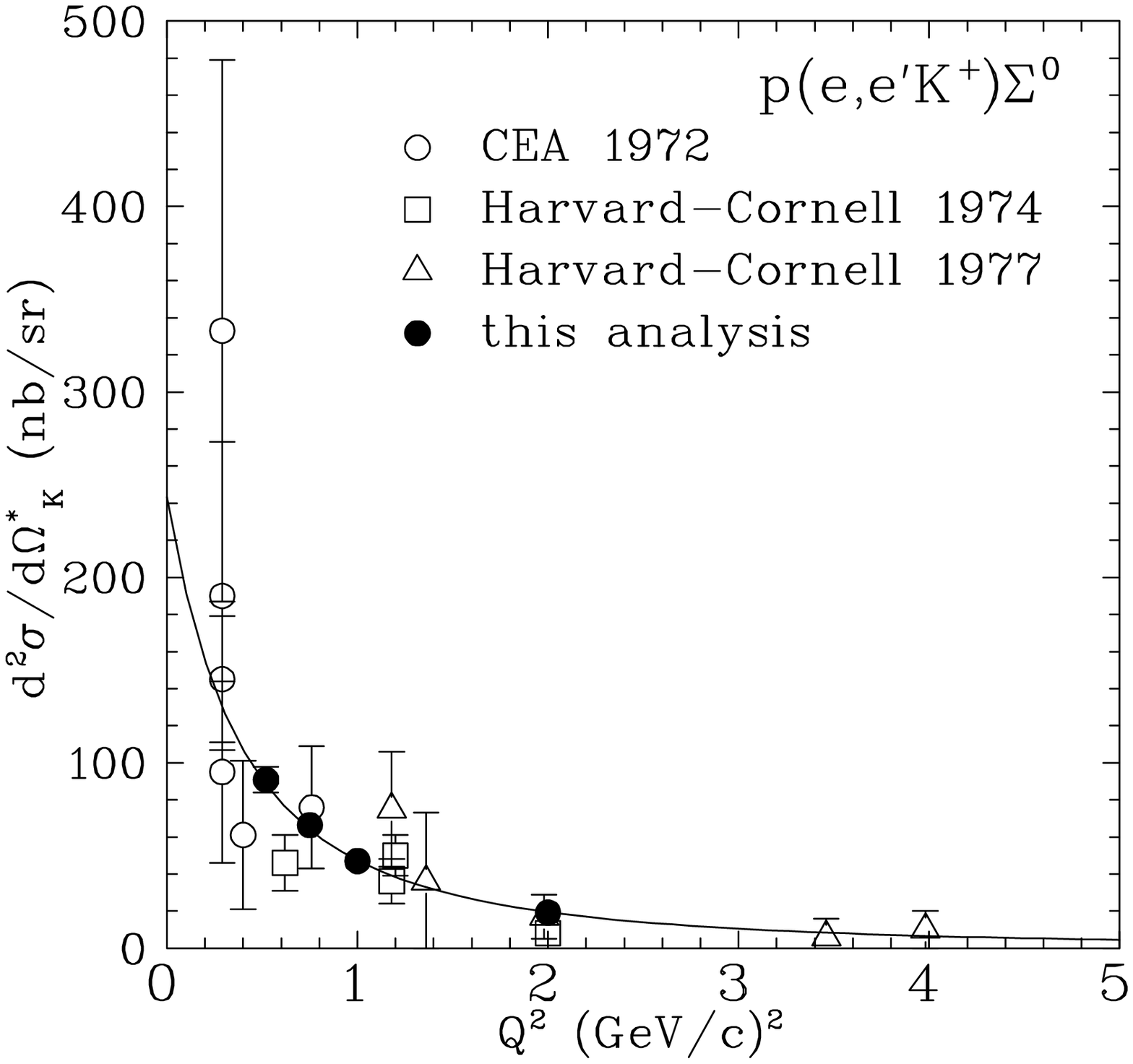}
\caption{Previous world data (see caption for
Fig.~\protect\ref{fig:world-lam-plus}) with the addition of the
highest-$\epsilon$ results of this analysis (solid points) scaled to 
$W$~=~2.15~GeV, $\avgthcm=0^\circ$, for the $p(e,e'K^+)\Sigo$
unseparated cross sections.}
\label{fig:world-sig-plus}
\end{center}
\end{figure}

The extracted p(e,e$'$K$^+$)$\Lambda$ and p(e,e$'$K$^+$)$\Sigo$
unseparated cross sections
are given in Table~\ref{tab:unsep}.  For the sake of comparison,
the unseparated cross sections at the highest $\epsilon$ values
(which were similar to those of the earlier data) are also shown in
Figs.~\ref{fig:world-lam-plus} and~\ref{fig:world-sig-plus} 
along with the previous
world data taken from \cite{bebek-prd0277,brownworlddata} (also,
see \cite{rmmthesis}). For the purposes of this plot, the E93-018
results have been scaled to $W$=2.15~GeV using the
parameterization in \cite{bebek-lt} (Equation \ref{eq:bebekw}) and
include a 5\% (6\%) scale error for the $\Lambda$ ($\Sigo$) data.  
The previous world data shown in this plot have been scaled to 
$W$~=~2.15~GeV and $\avgthcm=0^\circ$ using equations~\ref{eq:bebekw} 
and \ref{eq:tmin2}. It should be emphasized
that the data shown are at varying values of $t$, ranging from
0.05-3.0 GeV$^2$, so quantitative comparisons between data sets
should be performed with care. The $Q^2$ dependent parameterization in
Ref.~\cite{bebek-prd0277} and shown here is for data at 
$\avgthcm=0^\circ$. Qualitatively good agreement is seen with
previous data, and the new data do not significantly alter the
$Q^2$ parameterization derived from older data sets. 

% === TABLE ====================================
\begin{table}
\squeezetable
\begin{center}
\begin{tabular}{c c c c c}
\hline
& \textbf{$\Lambda$ channel} & & \\
\hline\hline
$\langle Q^2\rangle$ & $\langle W\rangle$ & $-t_{min}$~(GeV)$^2$
& $\epsilon$ & $\sigma_{\sss{T}} + \epsilon\sigma_{\sss{L}}$ \\
GeV$^2$ & GeV & & & nb/sr \\
\hline
0.52 & 1.84 & 0.22 & 0.552 & 367.6 $\pm$ 12.0 \\
 &  & & 0.771 & 391.5 $\pm$ 12.3 \\
 &  & & 0.865 & 405.3 $\pm$ 13.1 \\
\hline
0.75 & 1.84 & 0.30 & 0.462 & 329.7 $\pm$ 10.8 \\
 &  & & 0.724 & 357.4 $\pm$ 10.8 \\
 &  & & 0.834 & 381.1 $\pm$ 11.3 \\
\hline
1.00 & 1.81 & 0.41  & 0.380 & 293.9 $\pm$ 10.4 \\
&  & & 0.678 & 332.5 $\pm$ 11.3 \\
 & & & 0.810 & 340.3 $\pm$ 11.8 \\
\hline
2.00 & 1.84 & 0.74 & 0.363 & 184.5 $\pm$ 8.0 \\
&  & & 0.476 & 200.6 $\pm$ 7.0 \\
&  & & 0.613 & 202.9 $\pm$ 6.4 \\
\hline
& \textbf{$\Sigo$ channel} & & \\
\hline\hline
$\langle Q^2\rangle$ & $\langle W\rangle$ & $-t_{min}$~(GeV)$^2$
& $\epsilon$ & $\sigma_{\sss{T}} + \epsilon\sigma_{\sss{L}}$ \\
GeV$^2$ & GeV & & & nb/sr \\
\hline
0.52 & 1.84 & 0.31 & 0.545 & 75.4 $\pm$ 5.5 \\
 &  & & 0.757 & 87.3 $\pm$ 4.6 \\
 &  & & 0.851 & 86.2 $\pm$ 4.0 \\
\hline
0.75 & 1.84 & 0.41 & 0.456 & 54.2 $\pm$ 4.0 \\
 &  & & 0.709 & 64.7 $\pm$ 3.3 \\
 &  & & 0.822 & 63.0 $\pm$ 2.7 \\
\hline
1.00 & 1.81 & 0.55 & 0.375 & 37.9 $\pm$ 4.5 \\
 &  & & 0.663 & 43.6 $\pm$ 3.3 \\
 &  & & 0.792 & 42.4 $\pm$ 2.4 \\
\hline
2.00 & 1.84 & 0.95 & 0.352 & 17.0 $\pm$ 2.8 \\
 &  & & 0.461 & 16.2 $\pm$ 2.5 \\
 &  & & 0.598 & 18.3 $\pm$ 1.6 \\
\hline
\end{tabular}
\end{center}
\caption{Results for the unseparated $p(e,e'K^+)\Lambda$
and $p(e,e'K^+)\Sigo$  cross sections
used in the L/T separation: the uncertainties do not include 
the scale error of 5\% for the $\Lambda$ channel and 6\% for
the $\Sigo$ channel. Cross sections were extracted 
at $\avgthcm=0^\circ$ and at the $W$ and $Q^2$ values in the 
table, using the procedure outlined in section IV.C.  Note that because
a single experimental setting was used to acquire both $\Lambda$ and
$\Sigo$ data, the results are at different values of $-t_{min}$ for
fixed $\avgthcm$.}
\label{tab:unsep}
\end{table}
% ==============================================

% === GRAPHICS ====================================
% ==============================================

% === GRAPHICS ====================================
% ==============================================

\begin{figure} 
\begin{center}
\includegraphics[width=3.3in]{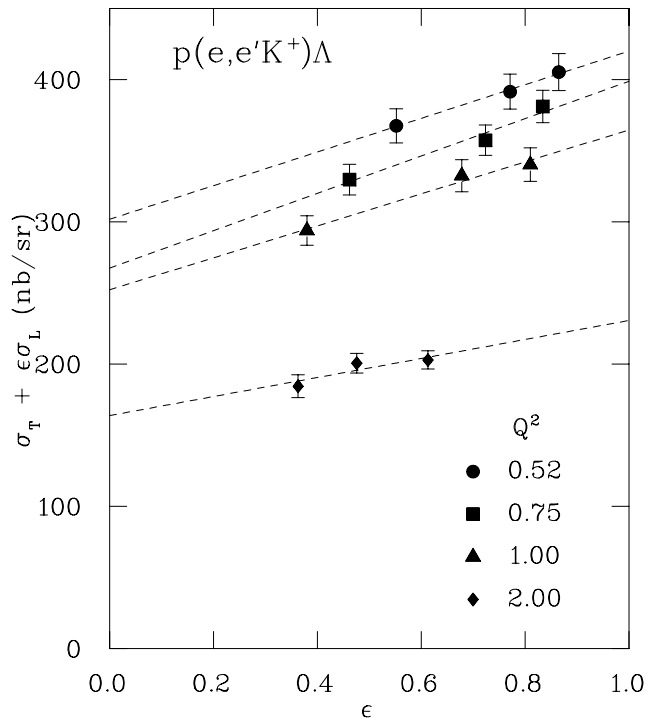}
\caption{Cross sections as a function of $\epsilon$ for the
$p(e,e'K^+)\Lambda$ process, shown with the linear fit to the data
that allows separation into the longitudinal and transverse
components.} 
\label{fig:lambda-lt}
\end{center}
\end{figure}

\begin{figure} 
\begin{center}
\includegraphics[width=3.3in]{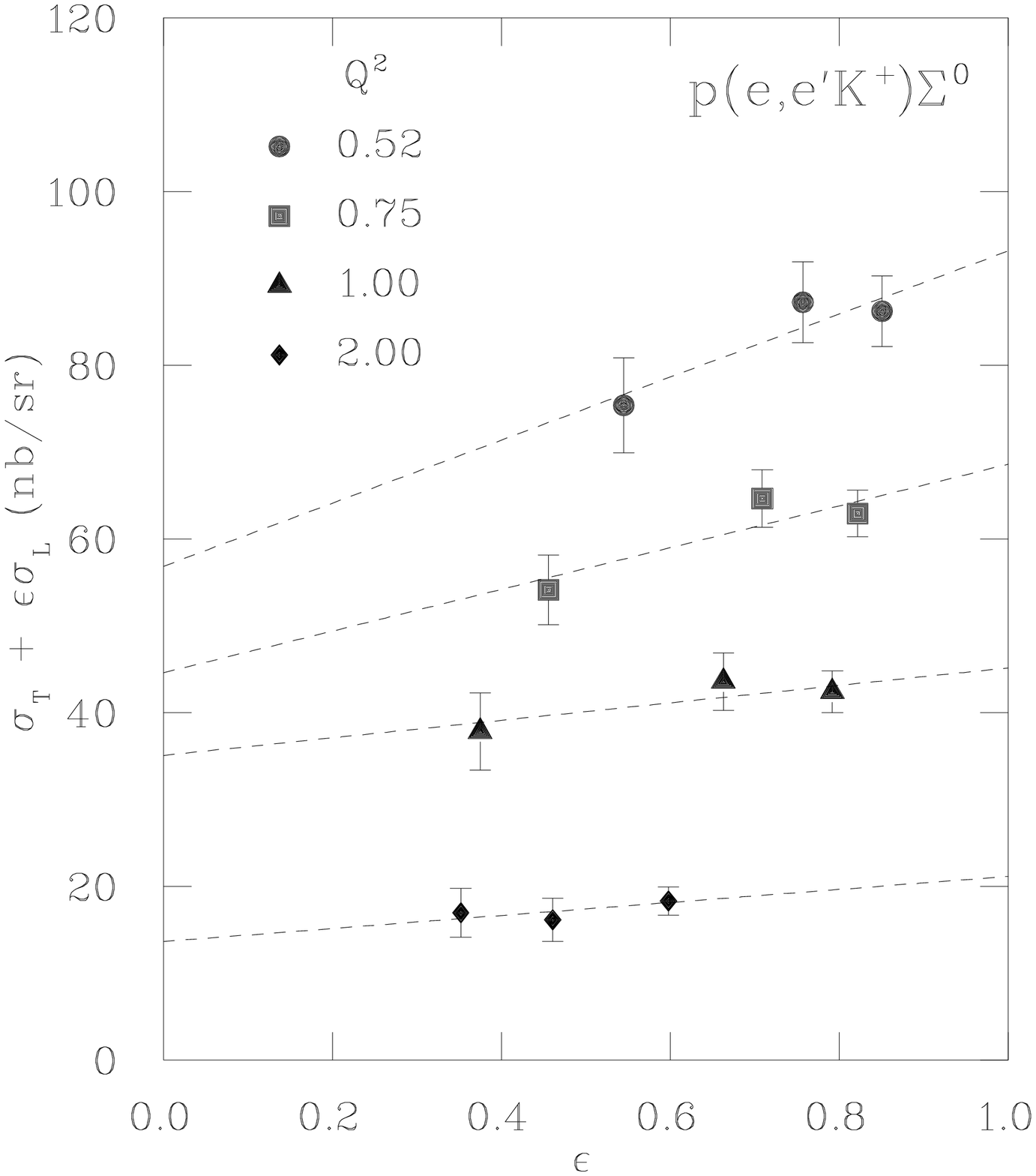}
\caption{Cross sections as a function of $\epsilon$ for the
$p(e,e'K^+)\Sigo$ process. The lines are the fit to the data,
allowing separation of the longitudinal and transverse
components. } 
\label{fig:sigma-lt}
\end{center}
\end{figure}

The unseparated cross sections are plotted as a function of $\epsilon$ in
Figures~\ref{fig:lambda-lt} and \ref{fig:sigma-lt}.  
A linear least-squares fit 
was performed at each value of $Q^2$ to determine the best straight line
($\sigma~=~\sigt~+~\epsilon\sigl$) through the points. The resulting values of
$\sigl$, $\sigt$, and $R$~=~$\sigl/\sigt$ are shown in
Table~\ref{tab:ltsep}.  Although only statistical and random systematic 
errors were used in the linear fit, the errors on the extracted values of
$\sigl$ and $\sigt$ include the scale errors added in quadrature
with the random errors. The quantity $R$ is
insensitive to scale errors.

\begin{figure} 
\begin{center}
\includegraphics[width=3.3in]{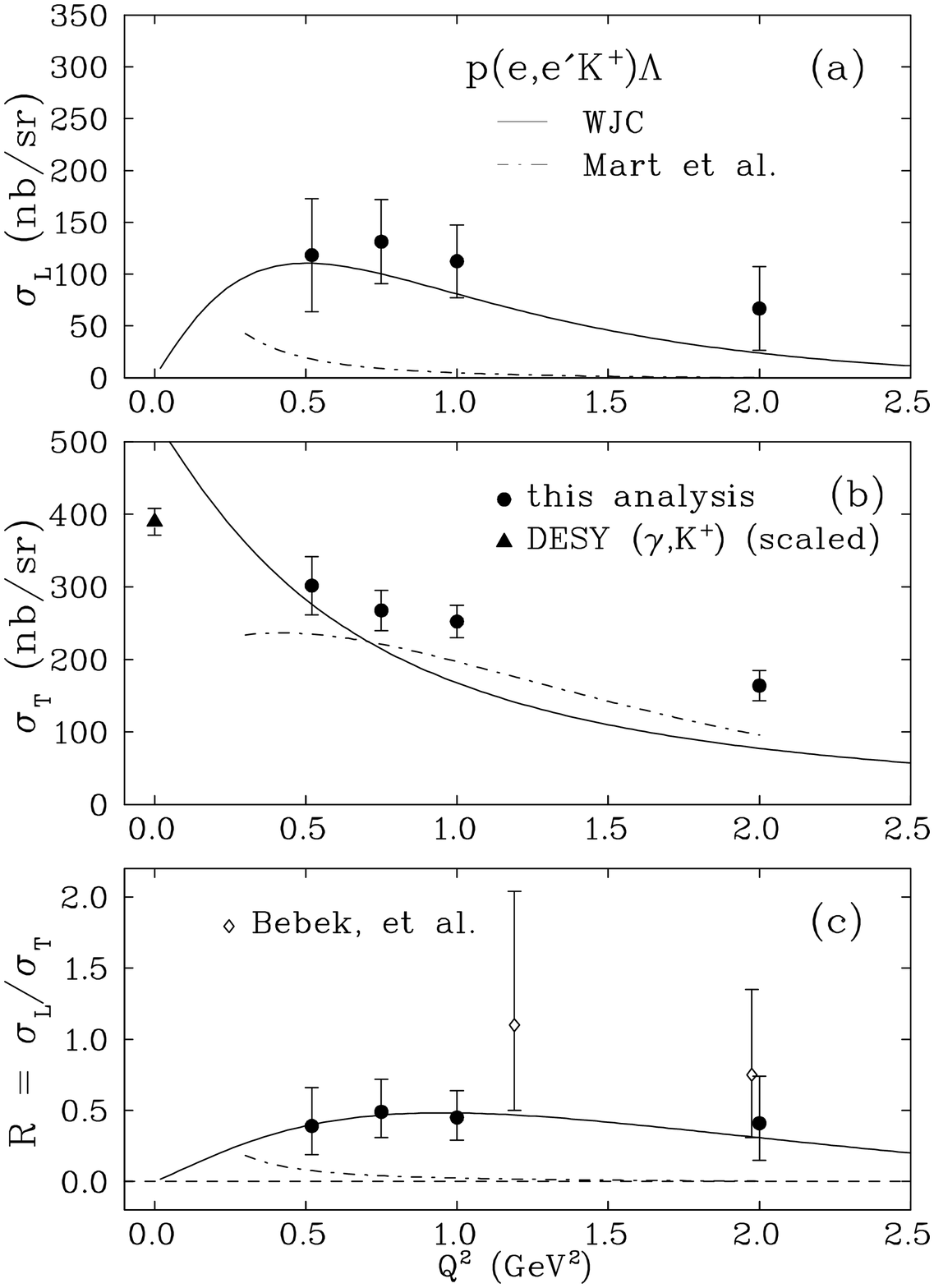}
\caption{Longitudinal (a) and transverse (b)
cross sections for $p(e,e'K^+)\Lambda$
as a function of $Q^2$. The ratio is shown in panel (c). The
data are compared to calculations of~\protect\cite{wjc1992} (solid line)
and~\protect\cite{Mar00} (dashed line). The open diamond
data are from~\protect\cite{bebek-lt}, and the solid diamond 
photoproduction data point is from~\protect\cite{feller}.}
\label{fig:lambda-t-sep}
\end{center}
\end{figure}

\begin{figure} 
\begin{center}
\includegraphics[width=3.3in]{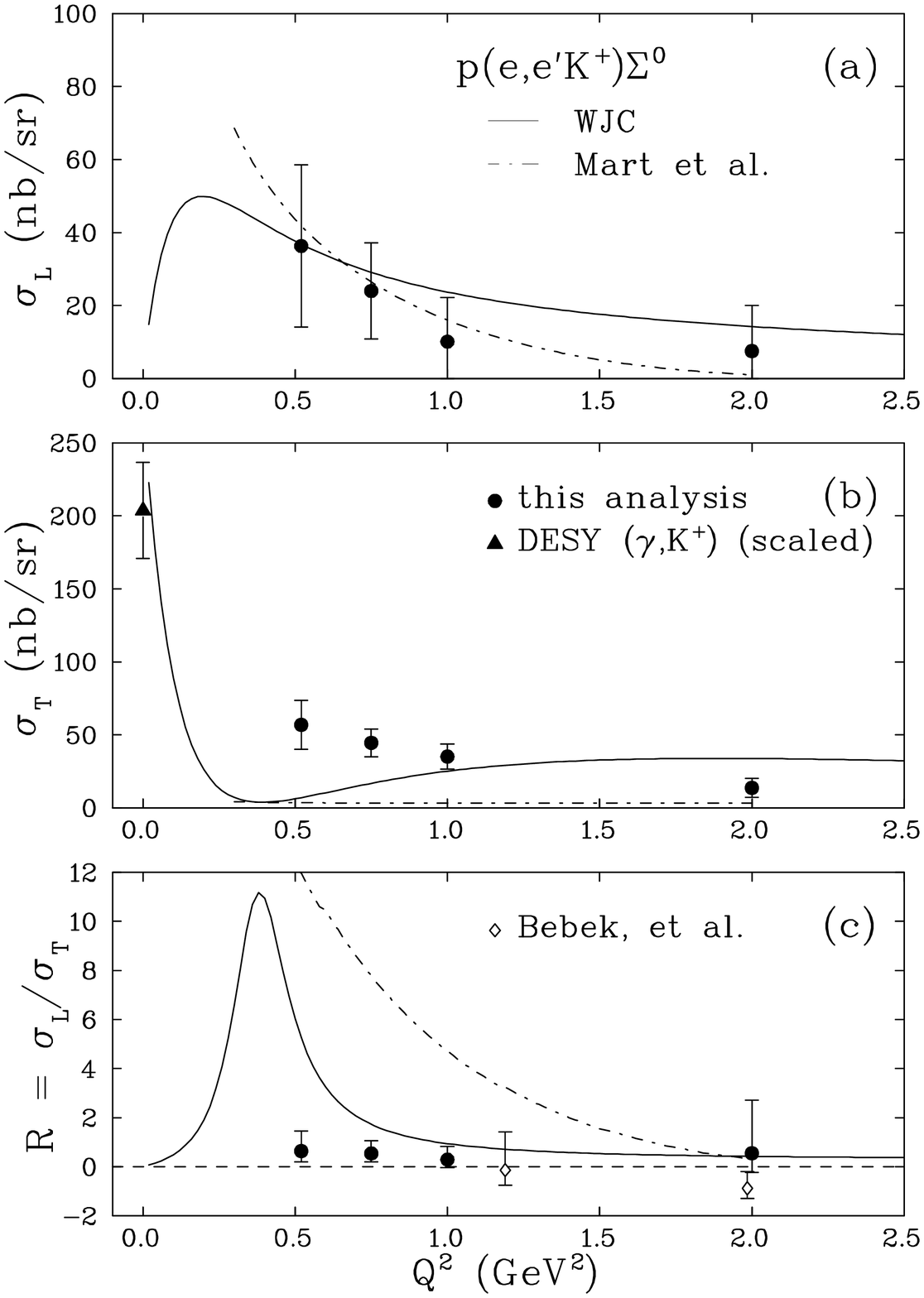}
\caption{Longitudinal (a) and transverse (b) cross
sections for $p(e,e'K^+)\Sigo$
as a function of $Q^2$. The ratio is shown in panel (c).
The calculations are again from~\protect\cite{wjc1992} (solid line)
and~\protect\cite{Mar00} (dashed line). The open diamond
data are from~\protect\cite{bebek-lt}, and the solid diamond 
photoproduction data point is from~\protect\cite{feller}.}
\label{fig:sigma-t-sep}
\end{center}
\end{figure}

The separated cross sections $\sigl$ and $\sigt$ for the $\Lambda$ channel
are plotted as a function of $Q^2$ in 
Figs.~\ref{fig:lambda-t-sep}(a) and \ref{fig:lambda-t-sep}(b),
respectively, along with other existing data. The equivalent plots
for the $\Sigo$ channel are in Fig.~\ref{fig:sigma-t-sep}.
Photoproduction points from~\cite{feller}
are also shown in the transverse components, taken at
comparable values of $W$ and $\thcm\sim 30^\circ$.
For these figures they are scaled to 
$W$~=~1.84~GeV, $\thcm = 0^\circ$ (corresponding to 
an upward adjustment of 1.3 (1.1) for $\Lambda$ ($\Sigo$)).
The third panel of each plot contains the ratio
$R=\sigl/\sigt$ as a function of $Q^2$, along with data
from~\cite{bebek-lt}.  The curves shown are from the WJC 
model~\cite{wjc1992} and from the unitary isobar model of 
Mart \textit{et al.} with its default parameterizations~\cite{Mar00}.

\begin{figure}
\begin{center}
\includegraphics[width=3.3in]{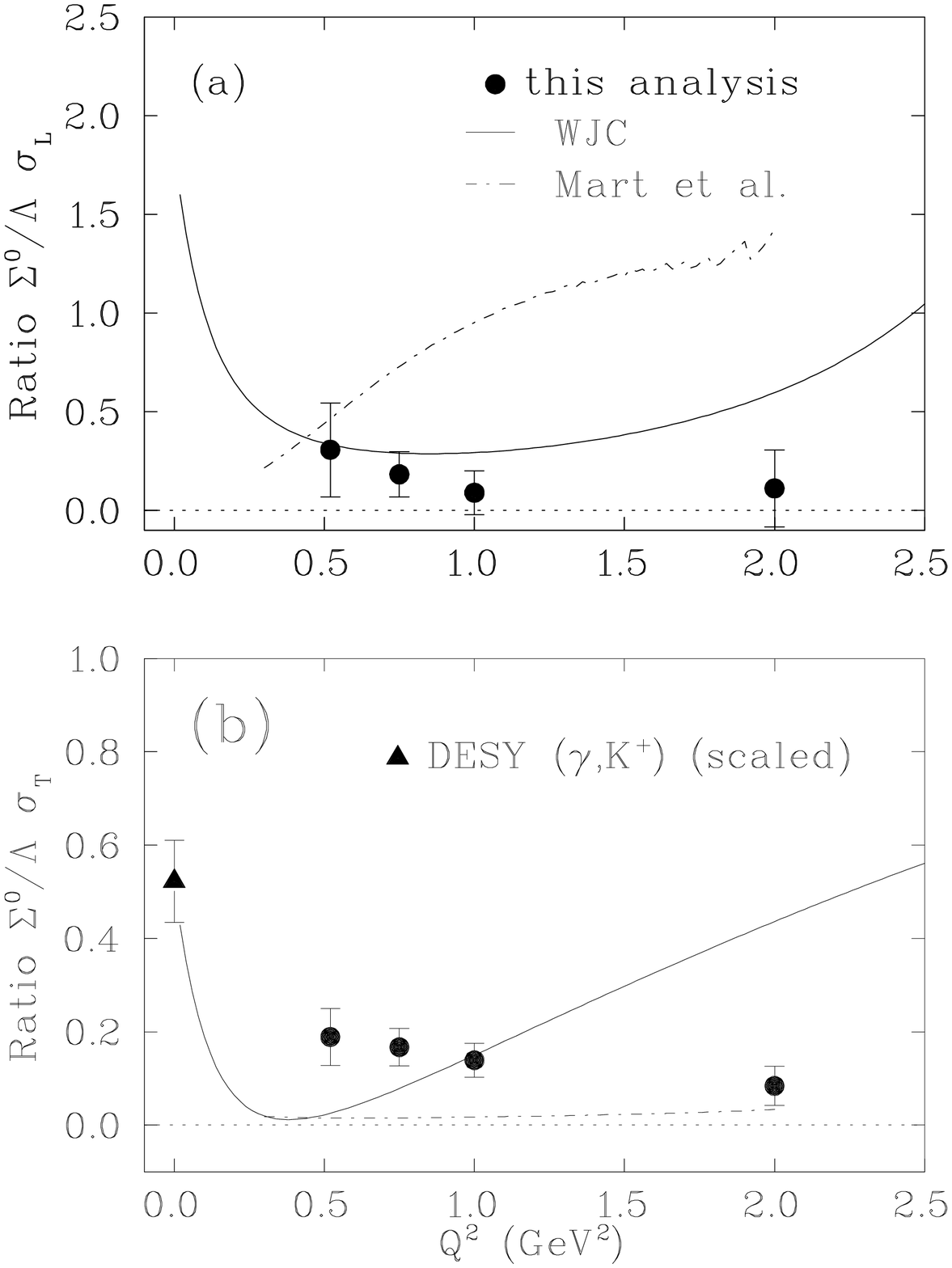}
\caption{Ratio of $\Sigo$/$\Lambda$ cross sections
as a function of $Q^2$, separated into longitudinal (a) and transverse (b)
 components, compared with the same two calculations as above.}
\label{fig:lamsig_wjc}
\end{center}
\end{figure}

Finally, the ratio of $\Sigo$/$\Lambda$ separated cross sections $\sigl$
and $\sigt$ are shown plotted vs.~$Q^2$ in
Figs.~\ref{fig:lamsig_wjc}(a) and \ref{fig:lamsig_wjc}(b), respectively,
along with curves from the two models.

% === TABLE ====================================
\begin{table}
\squeezetable
\begin{center}
\begin{tabular}{c c c c c}
\hline
& &\textbf{$\Lambda$ channel} & & \\
\hline\hline
$\langle Q^2\rangle$ & $\langle W\rangle$ & $\sigma_{\sss{L}}$ (nb/sr) &
$\sigma_{\sss{T}}$ (nb/sr) & R = $\sigma_{\sss{L}}/\sigma_{\sss{T}}$ \\
\hline
0.52 & 1.84 & 118.3 $\pm$ 54.6 & 301.8 $\pm$ 40.1 & 0.39$^{+0.27}_{-0.21}$ \\
0.75 & 1.84 & 131.3 $\pm$ 40.5 & 267.5 $\pm$ 27.8 & 0.49$^{+0.23}_{-0.18}$ \\
1.00 & 1.81 & 112.4 $\pm$ 35.2 & 252.3 $\pm$ 22.1 & 0.45$^{+0.19}_{-0.16}$ \\
2.00 & 1.84 &  66.8 $\pm$ 40.4 & 163.8 $\pm$ 20.7 & 0.41$^{+0.34}_{-0.26}$ \\
\hline
& &\textbf{$\Sigo$ channel} & & \\
\hline\hline
$\langle Q^2\rangle$ & $\langle W\rangle$ &  $\sigma_{\sss{L}}$ (nb/sr) &
$\sigma_{\sss{T}}$ (nb/sr) & R = $\sigma_{\sss{L}}/\sigma_{\sss{T}}$ \\
\hline
0.52 & 1.84 & 36.3 $\pm$ 22.2 & 56.9 $\pm$ 16.8 & 0.64$^{+0.81}_{-0.45}$ \\
0.75 & 1.84 & 24.0 $\pm$ 13.2 & 44.6 $\pm$  9.5 & 0.54$^{+0.52}_{-0.34}$ \\
1.00 & 1.81 & 10.1 $\pm$ 12.1 & 35.1 $\pm$  8.5 & 0.29$^{+0.54}_{-0.33}$ \\
2.00 & 1.84 &  7.5 $\pm$ 12.5 & 13.7 $\pm$  6.6 & 0.55$^{+2.15}_{-0.78}$ \\
\end{tabular}
\begin{tabular}{c c c c}
\hline
& \textbf{Ratio of $\Sigo$ / $\Lambda$} & & \\
\hline
$\langle Q^2\rangle$ & $\langle W\rangle$ &
$\sigma_{\sss{L}}(\Sigo)/\sigma_{\sss{L}}(\Lambda)$ &
$\sigma_{\sss{T}}(\Sigo)/\sigma_{\sss{T}}(\Lambda)$ \\
\hline
0.52 & 1.84 & 0.31 $\pm$ 0.24 & 0.19 $\pm$ 0.06 \\
0.75 & 1.84 & 0.18 $\pm$ 0.11 & 0.17 $\pm$ 0.04 \\
1.00 & 1.81 & 0.09 $\pm$ 0.11 & 0.14 $\pm$ 0.04 \\
2.00 & 1.84 & 0.11 $\pm$ 0.20 & 0.084 $\pm$ 0.042 \\
\hline
\end{tabular}
\end{center}
\caption{L/T separated cross section results from this analysis
for reactions p(e,e$'$K$^+$)$\Lambda$ and p(e,e$'$K$^+$)$\Sigo$,
and for the ratio of $\Sigo$ to $\Lambda$ cross sections.}
\label{tab:ltsep}
\end{table}
% ==============================================

% === GRAPHICS ====================================
% ==============================================
% === GRAPHICS ====================================
% ==============================================

It should be noted that the data for E93-018 were taken 
in parallel with another experiment in which angular distributions
of kaon electroproduction from hydrogen and deuterium were studied. 
In a few cases the kinematic settings were very similar, 
and comparisons were made with cross sections extracted from the 
analysis of~\cite{DMKthesis,jinseokthesis}.  
They are in excellent agreement (within 2.5\%),
when scaled to the same $Q^2$ and $W$ values 
using Eq.~\ref{eq:bebekq} and~\ref{eq:dmkw}. 

% === GRAPHICS ====================================
% ==============================================

% === GRAPHICS ====================================
% ==============================================

% === GRAPHICS ====================================
% ==============================================

\subsection{Comparison with Calculations}

As described in the introduction, several model calculations
of $\Lambda$ and $\Sigo$ electroproduction cross sections, 
using parameters fit to previous data, are
available. We have chosen to compare our data to the 
models in \cite{wjc1992} (WJC) and \cite{Mar00}, for
which calculations were readily available in the form
in which the data are presented here. The parameters
of each model were constrained by global fits to 
previously obtained unpolarized photo- and electroproduction data, 
and, through crossing arguments,
to kaon radiative capture.

For the $\Lambda$ channel, the WJC model reproduces reasonably
well the trends in both the longitudinal and transverse components
(Figs.~\ref{fig:lambda-t-sep}(a) and \ref{fig:lambda-t-sep}(b), 
respectively), although the transverse component is underpredicted.
The calculation of Ref.~\cite{Mar00}
qualitatively reproduces the transverse piece, which is constrained
by the photoproduction point, but not the longitudinal component.
One possible cause for the discrepancy could be the lack of 
knowledge of the
$Q^2$ dependence of the baryon form factors
entering in the $s$-channel~\cite{cotanchprivcomm}. 
%Very little data exist from which form factors for strange baryons 
%can be extracted. 
In their study of 
kaon electroproduction, David~\textit{et al.}, 
observed that $\sigl/\sigt$ was sensitively dependent on the choice
of baryon form factors, while rather insensitive to the
reaction mechanism~\cite{saclay-lyon}, whereas the unpolarized cross
section alone did not depend strongly on the baryon form factors.

For $\peek \Sigo$, the transverse component is underestimated by both
models and thus the ratio is overestimated
(see Fig.~\ref{fig:sigma-t-sep}). The strong peak in $R$ implied
by the WJC model is not observed in our data. The magnitude
of this peak in the WJC model is very sensitive to the CM energy, $W$, of 
the reaction, indicating that there are strong resonance 
contributions in the model. 
As in the case of the $\Lambda$ channel, it is likely that the form 
factors and the strengths of the various resonances entering the model could be
modified in order to give better agreement with the data.   In general,
models for the $\Sigo$ channel are harder to tune than 
for the $\Lambda$ because of the
influence of isovector $\Delta$ resonances (of spin 1/2 and 3/2 in the
model) in the $\Sigo$ channel and because of the lower quality/quantity
of available data. 

The ratio of the longitudinal cross sections for $\Sigo/\Lambda$ 
(Fig.~\ref{fig:lamsig_wjc}(a)) appears to mildly decrease
with increasing $Q^2$. This could arise, for example, from differences in
the behavior of the $g_{\sss \Sigma NK}(t)$ and 
$g_{\sss \Lambda NK}(t)$ form factors, if the longitudinal 
response is dominated by $t$-channel processes.  

We note that while Regge trajectory models are not expected to work well
at the rather low CM energies of our data, which are still within
the nucleon resonance region, our highest $Q^2$ results are in reasonable
agreement with the calculation of~\cite{vgl2}, both in the unseparated
cross sections and in the L/T components. At lower momentum transfer
our data indicate a larger longitudinal component to the
$\Sigo$ cross sections than predicted by their model, perhaps indicative
of the larger number of resonance contributions to the $\Sigo$ channel.

The ratio of the transverse cross sections for $\Sigo/\Lambda$ 
(Fig.~\ref{fig:lamsig_wjc}(b)) shows a mild decrease 
above $Q^2 \approx 0.52$~GeV$^2$. However, the inclusion of 
the DESY photoproduction data on the plot shows that there is
likely a rapid decrease in $R_T$ for $Q^2$ below 0.5~(GeV/c)$^2$.
This lower momentum region may be of interest for further study, 
particularly in the $\Sigo$ channel.

\subsection*{Conclusions}

Rosenbluth separated kaon electroproduction data in two hyperon
channels, $p\,(e,e'K^+)\Lambda$ and $p\,(e,e'K^+)\Sigo$, have been
presented.  These results are the most precise measurements of the separated
cross sections $\sigt$ and $\sigl$ made to date, particularly for the $\Sigo$
channel, and will help constrain theoretical models of these
electroproduction processes.  Such data allow access to 
baryon excitations that couple strongly to final states with 
strangeness but weakly to $\pi$-N systems. They also allow the 
possibility of mapping out the $Q^2$ evolution away from the 
photoproduction point, thereby providing a means
to extract electromagnetic form factors
and detailed information about the excited state wave functions. 
Used in conjunction with models they will allow one to learn more 
about the reaction dynamics of strangeness production.

\section*{Acknowledgements}
The Southeastern Universities Research Association (SURA) operates the
Thomas Jefferson National Accelerator Facility for the United States
Department of Energy under contract DE-AC05-84ER40150.
This work was also supported by the Department of Energy 
contract no.~W-31-109-ENG-38 (ANL) and by grants from the 
National Science Foundation. We also acknowledge informative 
discussions with C.~Bennhold.

% =========================================
% Finally, references

\end{document}